\documentclass[11pt,fleqn]{article}
\usepackage{epsfig}
\def\greaterthansquiggle{\raise.3ex\hbox{$>$\kern-.75em\lower1ex\hbox{$\sim$}}}
\def\lessthansquiggle{\raise.3ex\hbox{$<$\kern-.75em\lower1ex\hbox{$\sim$}}}
\newcommand{\grts}{\greaterthansquiggle}
\newcommand{\lets}{\lessthansquiggle}
\newcommand{\la}{\label}
\newcommand{\re}{\ref}
\newcommand{\ci}{\cite}
\newcommand{\beqn}{\begin{eqnarray}}
\newcommand{\eeqn}{\end{eqnarray}}
\newcommand{\bequ}{\begin{equation}}
\newcommand{\eequ}{\end{equation}}
\newcommand{\bsl}{\begin{sloppypar}}
\newcommand{\esl}{\end{sloppypar}}

\begin{document}
\null
\vspace{.1cm}\hfill DESY 00-098\\
\vspace{.1cm}\hfill UWThPh-2000-25\\
\vspace{.1cm}\hfill WUE-ITP-2000-016\\
\vspace{.1cm}\hfill HEPHY-PUB 731/2000\\
\vspace{.1cm}\hfill hep-ph/xxxxxxx\\
\vskip .4cm

\begin{center}
{\Large \bf Impact of $e^+$ and $e^-$ Beam Polarization\\[.4em]
on Chargino and Neutralino Production\\[.4em]
at a Linear Collider\\[.4em] 
}
\end{center}
\vskip 1.5em
{\large
{\sc G.~Moortgat--Pick$^{a}$\footnote{e-mail:
    gudrid@mail.desy.de},
A.~Bartl$^{b}$\footnote{e-mail:
     bartl@ap.univie.ac.at},  
H. Fraas$^{c}$\footnote{e-mail:
    fraas@physik.uni-wuerzburg.de},
W.~Majerotto$^{d}$\footnote{e-mail:
     majer@hephy.oeaw.ac.at.}%
}}\\[3ex]
{\footnotesize \it
$^{a}$ DESY, Deutsches Elektronen--Synchrotron, D--22603 Germany}\\
{\footnotesize \it
$^{b}$ Institut f\"ur Theoretische Physik, Universit\"at Wien, 
A--1090 Wien, Austria}\\
{\footnotesize \it
$^{c}$ Institut f\"ur Theoretische Physik, Universit\"at
W\"urzburg, D--97074 W\"urzburg, Germany}\\
{\footnotesize \it
$^{d}$ Institut f\"ur Hochenergiephysik der \"Osterreichischen
 Akademie der Wissenschaften,\\\phantom{$^{d}$} A--1050 Wien, Austria}
\vskip .5em
\par
\vskip .4cm

\begin{abstract}
We study the production processes 
$e^+ e^-\to \tilde{\chi}^+_i \tilde{\chi}^-_j$, $i,j=1,2$, and
$e^+ e^-\to \tilde{\chi}^0_m \tilde{\chi}^0_n$, $m,n=1,\ldots,4$,  
working out the advantages of polarizing both beams. 
For $e^+ e^-\to \tilde{\chi}^+_1 \tilde{\chi}^-_1$ with
$\tilde{\chi}^-_1\to \tilde{\chi}^0_1 e^- \bar{\nu}$ and
$e^+ e^-\to \tilde{\chi}^0_1 \tilde{\chi}^0_2$ with 
$\tilde{\chi}^0_2 \to \tilde{\chi}^0_1 e^+ e^-$ we perform 
a detailed analysis, including the complete spin correlations between 
production and decay. We analyze 
the forward--backward asymmetry of the decay electron for various beam 
polarizations. We also study polarization asymmetries in 
$e^+ e^-\to \tilde{\chi}^0_1 \tilde{\chi}^0_2$. These asymmetries strongly
constrain the gaugino parameter $M_1$
and the masses $m_{\tilde{e}_L}$, $m_{\tilde{e}_R}$, $m_{\tilde{\nu}}$
also if $m_{\tilde{e},\tilde{\nu}}\ge \sqrt{s}/2$.
We give numerical predictions for three scenarios for a linear 
collider with $\sqrt{s}=500-1000$~GeV.
\end{abstract}
\vspace{1em}
\hfill
\section{Introduction}
The search for supersymmetric (SUSY) particles and the determination of their
properties will be one of the main goals of a future $e^+e^-$ linear collider.
Particularly interesting will be the experimental study of charginos and
neutralinos, 
which are the quantum mechanical mixtures of the charged and 
neutral gauginos and
higgsinos, the SUSY partners of the charged and
neutral gauge and Higgs bosons. In the
Minimal Supersymmetric Standard Model (MSSM) there are 
two charginos $\tilde{\chi}^{\pm}_i$, $i=1,2$, and
four neutralinos $\tilde{\chi}_i^0$, $i =1,\ldots,4$. 
 Usually the lightest neutralino $\tilde{\chi}_1^0$ is
the lightest SUSY particle LSP. 
The masses and couplings of the charginos are determined by the SUSY 
parameters 
$M_2$, $\mu$ and $\tan\beta$. The neutralino properties depend in addition on 
the gaugino mass parameter $M_1$.
An $e^+ e^-$ linear collider, where charginos and neutralinos can be pair 
produced, will allow a precise determination of the SUSY parameters involved.

Previous papers mainly analyzed production cross sections and decay
branching ratios in the MSSM (see, e.g. \ci{bartl,Ambrosanio} and 
references therein). 
Recently a method for determining the SUSY parameters $M_2$, $\mu$ and 
$\tan\beta$ by measuring suitable observables in chargino production
$e^+ e^-\to \tilde{\chi}^+_i \tilde{\chi}^-_j$ ($i,j=1,2$) has been proposed
\ci{Choi}. The gaugino mass parameter $M_1$ can in principle
be determined from the neutralino mass spectrum \ci{Choi,Kneur}.
Models with an extended neutralino sector have been discussed in 
\ci{Gudi_sitges, HesselFranke}.
A detailed study of the neutralino system is also helpful for examining
the question whether the MSSM or another SUSY model is realized in nature. 

In the present paper we study the production and decay of
charginos and neutralinos with both the $e^-$ 
and the $e^+$ beam polarized. 
We show that suitably polarizing both beams has three advantages: 
One can gain higher cross sections
and thereby reduce the experimental errors.
By measuring suitable observables one can get additional
information on the mixing components of charginos and neutralinos as well as 
on the masses of the exchanged $\tilde{\nu}_{e}$, $\tilde{e}_L$, 
$\tilde{e}_R$. Moreover, the background can be reduced by appropriately
polarizing the beams.

In the calculation of the decay angular distributions
one has to take into account the spin correlations between
production and decay of the charginos and 
neutralinos. They are particularly important near 
threshold. 
The processes $e^+ e^-\to \tilde{\chi}^+_i \tilde{\chi}^-_j$, 
$\tilde{\chi}^+_i\to \tilde{\chi}^0_k \ell^+ \nu$ and
$e^+ e^-\to \tilde{\chi}^0_i \tilde{\chi}^0_j$, 
$\tilde{\chi}^0_i \to \tilde{\chi}^0_k \ell^+ \ell^-$,
including the full spin correlations, have been studied in 
\ci{Gudi_char,Gudi_physrev}.
In \ci{Gudi_neut, Gudi_diss} 
we have given the complete analytical formulae for longitudinally
$e^-$ and $e^+$ polarized beams in the laboratory system. 

In our numerical analysis 
we will consider three
scenarios, where $\tilde{\chi}^0_{1,2}$, $\tilde{\chi}^{\pm}_1$ 
are gaugino--like and
which differ in $\tan\beta$ and the selectron masses.
We analyze 
the $e^-$ and $e^+$
polarization dependence of all production sections accessible at 
a linear collider in the 500-1000 GeV range.
In the case of $\tilde{\chi}^+_1 \tilde{\chi}^-_1$ pair
production and associated 
$\tilde{\chi}^0_1 \tilde{\chi}^0_2$ production we study the dependence 
of the cross section and the forward--backward asymmetry of the decay electron 
on the beam polarizations and on the 
masses of the exchanged $\tilde{\nu}$, 
$\tilde{e}_L$ and $\tilde{e}_R$. We also relax the GUT 
relation between $M_1$ and $M_2$,  $M_1/M_2 = \frac{5}{3} \tan^2\Theta_W$,
and study the $M_1$ dependence of
the forward-backward asymmetry of the decay electron 
and the polarization asymmetry. 
\section{Spin Correlations between Production and Decay}\la{sec:1}
\mathindent0cm
\bsl
The
helicity amplitudes for the production processes 
\beqn
&&\quad e^+ e^-\to \tilde{\chi}^+_i \tilde{\chi}^-_j, 
\la{eq_1a}\\
&&\quad
e^+ e^-\to \tilde{\chi}^0_i \tilde{\chi}^0_j,
\la{eq_1b}
\eeqn
are denoted by $T_P^{\lambda_i\lambda_j}$ 
and those for the decay processes
\beqn
&&\quad
\tilde{\chi}^{+}_{i}\to \tilde{\chi}^0_{k} \ell^{+}
\stackrel{(-)}{\nu_{\ell}},
\la{eq_2a}\\
&&\quad
\tilde{\chi}^{0}_{i}\to \tilde{\chi}^0_{k} \ell^+ \ell^-
\la{eq_2b}
\eeqn
by $T_{D,\lambda_i}$ (and analogously $T_{D,\lambda_j}$
for the decaying $\tilde{\chi}^-_j$ and $\tilde{\chi}^0_j$).
The corresponding Feynman diagrams are given in Figs.~1a 
and b. 
\esl

The amplitude 
squared of the combined process of production and decay is
(summed over helicities):
\bequ
|T|^2=|\Delta(\tilde{\chi}_i)|^2 |\Delta(\tilde{\chi}_j)|^2 
\rho_P^{\lambda_i \lambda_j \lambda^{'}_i \lambda^{'}_j} 
\rho_{D, \lambda^{'}_i\lambda_i}\rho_{D,\lambda_j^{'} \lambda_j}.
\la{eq4_4e}
\eequ
It is composed of the (unnormalized) spin density production matrix
\bequ
\rho_P^{\lambda_i\lambda_j\lambda_i^{'}\lambda_j^{'}}=T_P^{\lambda_i\lambda_j}
T_P^{\lambda_i^{'}\lambda_j^{'}*},
\la{eq4_4f}
\eequ
the decay matrices
\bequ
\rho_{D, \lambda_i^{'} \lambda_i}=T_{D, \lambda_i} T_{D, \lambda_i^{'}}^{*},
\quad
\rho_{D, \lambda_j^{'} \lambda_j}=T_{D,\lambda_j} T_{D,\lambda_j^{'}}^*,
\la{eq4_4g}
\eequ
and the propagator 
\bequ
\Delta(\tilde{\chi}_{k})=1/[p^2_k-m_{k}^2+i m_{k} \Gamma_{k}].
\eequ
Here $p_k^2$, $\lambda_{k}$, $m_{k}$ and $\Gamma_{k}$ denote the 
four--momentum squared, helicity, mass and total width of 
$\tilde{\chi}_{k}$. For this propagator we use the narrow--width 
approximation. 

\bsl
The density matrices can be expanded in terms of the Pauli matrices 
\nolinebreak$\sigma^a$: \esl
\beqn 
\rho_P^{\lambda_i\lambda_j\lambda_i^{'}\lambda_j^{'}}&=&
(\delta_{\lambda_i\lambda_i^{'}} \delta_{\lambda_j\lambda_j^{'}}
P(\tilde{\chi}_i \tilde{\chi}_j)
+\delta_{\lambda_j\lambda_j^{'}}\sum_{a=1}^3
\sigma^a_{\lambda_i\lambda_i^{'}}\Sigma^a_P(\tilde{\chi}_i)
\nonumber\\ &&
+\delta_{\lambda_i\lambda_i^{'}}\sum_{b=1}^3
\sigma^b_{\lambda_j\lambda_j^{'}}\Sigma^b_P(\tilde{\chi}_j)
+\sum_{a,b=1}^3\sigma^a_{\lambda_i\lambda_i^{'}}
\sigma^b_{\lambda_j\lambda_j^{'}}
\Sigma^{ab}_P(\tilde{\chi}_i \tilde{\chi}_j)),\la{eq4_4h}\\
\rho_{D,\lambda_i^{'}\lambda_i}&=&(\delta_{\lambda_i^{'}\lambda_i}
D(\tilde{\chi}_i)+\sum_{a=1}^3
\sigma^a_{\lambda_i^{'}\lambda_i} \Sigma^a_D(\tilde{\chi}_i)),\la{eq4_4i}\\
\rho_{D,\lambda_j^{'}\lambda_j}&=&(\delta_{\lambda_j^{'}\lambda_j}
D(\tilde{\chi}_j)+\sum_{b=1}^3
\sigma^b_{\lambda_j^{'}\lambda_j} \Sigma^b_D(\tilde{\chi}_j)).\la{eq4_4j}
\eeqn
We have chosen the polarization vectors such that $\Sigma^1_P(\tilde{\chi}_{i,j})$
 describes the 
transverse polarization in the production plane, 
$\Sigma^2_P(\tilde{\chi}_{i,j})$ denotes the 
polarization perpendicular to the production plane and 
$\Sigma^3_P(\tilde{\chi}_{i,j})$ describes 
the longitudinal polarization of the chargino or neutralino, respectively. 
$\Sigma^{ab}_P(\tilde{\chi}_i\tilde{\chi}_j)$ 
is due to correlations between 
the polarizations of both charginos or neutralinos, respectively. 
The complete analytical expressions for 
the production density matrix and for the decay matrices are given in
\ci{Gudi_char,Gudi_neut}.

The amplitude squared $|T|^2$ 
of the combined process of production and decay, eq.~(\re{eq4_4e}), can
be rewritten as:
\beqn
&&\mbox{\hspace{-.7cm}}
|T|^2=4|\Delta(\tilde{\chi}_i)|^2|
\Delta(\tilde{\chi}_j)|^2
         \Big(P(\tilde{\chi}_i \tilde{\chi}_j) 
D(\tilde{\chi}_i) D(\tilde{\chi}_j)
    +\sum^3_{a=1}\Sigma_P^a(\tilde{\chi}_i) 
\Sigma_D^a(\tilde{\chi}_i) 
D(\tilde{\chi}_j)\nonumber\\
& &\phantom{4|}
+\sum^3_{b=1}\Sigma_P^b(\tilde{\chi}_j) \Sigma_D^b(\tilde{\chi}_j)
D(\tilde{\chi}_i)
    +\sum^3_{a,b=1}\Sigma_P^{ab}(\tilde{\chi}_i\tilde{\chi}_j)
 \Sigma^a_D(\tilde{\chi}_i) 
\Sigma^b_D(\tilde{\chi}_j)\Big).
\la{eq4_5}
\eeqn
The differential cross section is then given by
\begin{equation}
d\sigma_e=\frac{1}{2 s}|T|^2 (2\pi)^4
\delta^4(p_1+p_2-\sum_{i} p_i) d{\rm lips}(p_3\ldots p_{10})\label{eq_13},
\end{equation}
where $d{\rm lips}(p_3,\ldots,p_{10})$ 
is the Lorentz invariant phase space element.

If one neglects all spin correlations between production and decay only the 
first term in (\re{eq4_5}) contributes. The second and third term in 
(\re{eq4_5}) describe the spin correlations between the production and the 
decay process. The last term is due to spin--spin correlations between 
both decaying charginos or neutralinos.
\section{Numerical Analysis and Discussion}\la{sec:3}
In the following analysis we use the MSSM \ci{Haber-Kane} as our general 
framework. 
The masses and couplings of neutralinos and charginos are determined
 by the 
parameters $M_1$, $M_2$, $\mu$, $\tan\beta$, which can be chosen real if CP 
violation is neglected. Moreover, one usually makes use of the GUT relation 
\bequ
M_1=\frac{5}{3} M_2 \tan^2\Theta_W.
\la{eq_m1}
\eequ
The explicit expressions for the neutralino and chargino mass mixing matrices
can be found in \ci{bartl} (note that in 
Refs.~\ci{bartl,Gudi_char} the notation 
$M'$ and $M$ for $M_1$ and $M_2$ was used).

In our analysis we study chargino and neutralino production and decays 
in three scenarios, which we denote by A1, A2 and B. 
The corresponding parameters are given in 
Table~\ref{tab:1}.
In scenario~A1 one has
$\tan\beta=3$ and the masses  
$m_{\tilde{e}_L}=176$~GeV, $m_{\tilde{e}_R}=132$~GeV, 
$m_{\tilde{\nu}}=161$~GeV \ci{Blair}.
In this scenario $\tilde{\chi}^0_1$ is $\tilde{B}$--like, $\tilde{\chi}^0_2$
 is $\tilde{W}^3$--like, and $\tilde{\chi}^{\pm}_1$ is 
$\tilde{W}^{\pm}$--like and $\tilde{\chi}^{\pm}_2$  $\tilde{H}^{\pm}$--like. 
Scenario~A2 differs from A1 only by the higher $\tilde{e}_L$ mass 
$m_{\tilde{e}_L}=500$~GeV. 
In scenario~B one has $\tan\beta=30$ and 
$m_{\tilde{e}_L}=217$~GeV, $m_{\tilde{e}_R}=183$~GeV and 
$m_{\tilde{\nu}}=202$~GeV \ci{Blair}. 
In this scenario $\tilde{\chi}^0_1$, 
$\tilde{\chi}^0_2$ and $\tilde{\chi}^{\pm}_1$ are also gaugino--like.
\subsection{Beam Polarization Effects in Chargino Production}
\la{sec:31}
\bsl
We study the dependence of the cross sections 
$\sigma(e^+ e^-\to\tilde{\chi}^+_i \tilde{\chi}^-_j)$, $i,j=1,2$, 
on the electron beam 
polarization $P_{e^-}$ and positron beam polarization $P_{e^+}$ (with
$P_{e^\pm}=\{-1, 0, 1\}$ for $\{$left--,un--, right--$\}$ polarized)
at $\sqrt{s}=m_{\tilde{\chi}^{+}_i}+m_{\tilde{\chi}^-_j}+10$~GeV for 
scenarios~A1 and B. We choose $\sqrt{s}$ not too far from threshold, 
because the spin correlations to be discussed below
are largest near threshold \ci{Gudi_char}.
As an abbrevation we write in the following just the final state 
$\tilde{\chi}^+_i \tilde{\chi}^-_j$ for the process
$e^+ e^-\to \tilde{\chi}^+_i \tilde{\chi}^-_j$.
\esl 
\begin{itemize}
\item $\tilde{\chi}^+_1 \tilde{\chi}^-_1$\\
{\it Scenario~A1} (see Fig.~\re{fig_2}a): 
Due to the $\tilde{\nu}_e$ exchange and
the gaugino character of $\tilde{\chi}^{\pm}_1$ 
a left polarized electron beam and a right polarized positron beam 
lead to the largest cross section.
For $P_{e^-}=-85\%$ and $P_{e^+}=+60\%$ 
we obtain 410~fb, which means that the cross section 
is enhanced by a factor of 
about three with respect to unpolarized beams (Table~\re{table_11}).

{\it Scenario~B}: One gets a similar dependence on the beam polarizations 
and the same enhancement factor as in 
scenario~A1. With $P_{e^-}=-85\%$ and $P_{e^+}=+60\%$ a cross section
of 709~fb is reached (Table~\re{table_11}).
\item $\tilde{\chi}^{\pm}_1 \tilde{\chi}^{\mp}_2$\\ 
{\it Scenario~A1} (see Fig.~\re{fig_2}b): 
In this case $\gamma$ exchange does not contribute and 
$\tilde{\nu}_e$ exchange is suppressed due to the higgsino--like
$\tilde{\chi}^-_2$. 
Therefore, right polarized electrons and left polarized positrons are 
favoured. Due to the different character of $\tilde{\chi}^+_1$ and
$\tilde{\chi}^-_2$ the cross section is small and reaches only 
6~fb  with $P_{e^-}=+85\%$ and $P_{e^+}=-60\%$ (Table~\re{table_11}).

{\it Scenario~B} (see Fig.~\re{fig_2}d):
In this case somewhat higher cross sections of about 10~fb
can be reached with
 $P_{e^-}=+85\%$ and $P_{e^+}=-60\%$.
\item $\tilde{\chi}^{+}_2 \tilde{\chi}^{-}_2$\\
{\it Scenario~A1} (see Fig.~\re{fig_2}c):
As $\tilde{\chi}^{\pm}_2$ is higgsino--like $\tilde{\nu}_e$ exchange is 
negligible. The cross section is enhanced if the electrons are left and the 
positrons are right polarized, mainly due to the $\gamma-Z^0$ interference. 
For $P_{e^-}=-85\%$ and $P_{e^+}=+60\%$
it is enhanced by a factor 2.5 and reaches 113~fb.

{\it Scenario~B}:
One gets a similar dependence on the beam polarizations
and the same enhancement factor as in 
scenario~A1. With $P_{e^-}=-85\%$ and $P_{e^+}=+60\%$ a cross section
of 194~fb is reached (Table~\re{table_11}). 
\end{itemize}
A summary of our results for scenario~A1 and B is given in
Table~\re{table_11}.

Since $\tilde{\nu}_e$ exchange  
favours left polarized electron beams and right polarized
positron beams, one expects for gaugino--like scenarios 
the following sequence of polarized cross sections \ci{Gudi_diss,LC99_char} 
(for 
$|P_{e^-}|=85$\% and $|P_{e^+}|=60$\%): 
\bequ
\sigma^{-+}>\sigma^{-0}>\sigma^{00}>\sigma^{--}>\sigma^{++}>
\sigma^{+0}>\sigma^{+-}.
\la{eq_gauge}
\eequ
Here $(-+)$ etc. denotes the sign of the electron polarization $P_{e^-}$ and 
of the positron polarization $P_{e^+}$, respectively.
 
On the other hand, 
for pure higgsinos one would have due to $Z^0$ exchange
\bequ
\sigma^{-+}>\sigma^{+-}>\sigma^{-0}>
\sigma^{00}>\sigma^{+0}>\sigma^{--}>\sigma^{++}.
\la{eq_higgs}
\eequ
These orderings are also 
valid if the decays are included. 

The relations
(\re{eq_gauge}), (\re{eq_higgs}) are, however, modified by 
the $\gamma$ exchange contribution.
Nevertheless, one can get additional information by using
polarized electron {\it and} positron beams, because the sequences of 
polarized cross sections for gaugino--like and higgsino--like 
scenarios are different. 
If only the electron beam is polarized, one would 
obtain in both scenarios the same sequence of polarized cross sections, namely 
$\sigma^{-0}>\sigma^{00}>\sigma^{+0}$.

In scenarios~A1 and ~B we get the results in Table~\re{table_11}
for $|P_{e^-}|=85\%$ and $|P_{e^+}|=60\%$
at $\sqrt{s}=m_{\tilde{\chi}^{\pm}_i}+m_{\tilde{\chi}^{\pm}_j}+10$~GeV.
For $e^+ e^-\to \tilde{\chi}^+_1 \tilde{\chi}^-_1$ 
the relation (\re{eq_gauge}) is fulfilled.
However, for $e^+ e^-\to \tilde{\chi}^{+}_2 \tilde{\chi}^-_2$
(higgsino--like $\tilde{\chi}^{\pm}_2$)
the sequence
is different from relation (\re{eq_higgs}) 
due to $\gamma$ exchange and in particular  
$\gamma Z^0$ interference. 

Usually, one defines an effective polarization
\bequ
P_{eff}=\frac{P_{e^-}-P_{e^+}}{1-P_{e^-} P_{e^+}}\quad.\la{eq_eff}
\eequ
The error of the effective polarization
can be reduced when both beams are polarized. 
\subsection{Beam Polarization Effects in Neutralino Production}
\la{sec:32}
In this subsection we study the dependence of
the cross section $\sigma(e^+ e^-\to \tilde{\chi}^0_i \tilde{\chi}^0_j)$,
$i, j=1,\ldots,4$, 
on the longitudinal beam polarizations 
$P_{e^-}$ and $P_{e^+}$ 
at $\sqrt{s}=(m_{\tilde{\chi}^0_i}+m_{\tilde{\chi}^0_j})+30$~GeV, 
for scenario~A1, A2 and B. In the following we again denote
the production 
process $e^+e^-\to \tilde{\chi}^0_i \tilde{\chi}^0_j$ 
by its final state $\tilde{\chi}^0_i \tilde{\chi}^0_j$.
\begin{itemize}
\item $\tilde{\chi}^0_1 \tilde{\chi}^0_2$\\ 
{\it Scenario~A1} (see Fig.~\re{fig_3}a):
Since $\tilde{\chi}^0_1$ 
is mostly $\tilde{B}$--like and $\tilde{\chi}^0_2$ is mostly
$\tilde{W}^3$--like mainly $\tilde{e}_L$ and $\tilde{e}_R$ exchange in 
the t-- and u-- channel contribute.
One gets the largest cross section for left polarized 
electrons and right polarized positrons. Due to the strong $\tilde{B}$ 
component of $\tilde{\chi}^0_1$
one gets a slight enhancement also for right polarized electrons 
and left polarized positrons. For $P_{e^-}=-85\%$ and $P_{e^+}=+60\%$
the cross section goes up to 56~fb, which is larger by a factor of 1.8
than for unpolarized beams.

{\it Scenario~A2} (see Fig.~\re{fig_3}b):
Since $m_{\tilde{e}_L} \gg m_{\tilde{e}_R}$
$\tilde{e}_L$ exchange is suppressed leading to an
enhancement for right polarized electrons and left polarized 
positrons. With $P_{e^-}=+85\%$ and $P_{e^+}=-60\%$ one reaches 
35~fb, that is an enhancement by a factor 
2.7 with respect to unpolarized beams.

{\it Scenario~B} (see Fig.~\re{fig_3}c):
Due to the different couplings for large $\tan\beta$
the dependence on the beam 
polarizations slightly changes. 
For $P_{e^-}=-85\%$ and $P_{e^+}=+60\%$ one has an enhancement
factor of 2.4 with respect to the 
unpolarized case. The cross section reaches 50~fb. 
\item $\tilde{\chi}^0_2 \tilde{\chi}^0_2$\\ 
{\it Scenario~A1} (see Fig.~\re{fig_3}d):
Owing to the $\tilde{W}^3$ nature of $\tilde{\chi}^0_2$, the 
cross section is governed by $\tilde{e}_L$ exchange. Hence the cross 
section is largest for left polarized 
electrons and right polarized positrons. It reaches 121~fb for
$P_{e^-}=-85\%$ and $P_{e^+}=+60\%$, which gives an enhancement by a factor 
of 3 with respect to unpolarized beams. 

{\it Scenario~B}:
One gets the same enhancement factor compared to the unpolarized case
as in scenario~A1. Due to the smaller 
lepton-slepton-neutralino couplings the cross sections are about a factor 
two smaller. For
$P_{e^-}=-85\%$ and $P_{e^+}=+60\%$ one gets 59~fb.
\item $\tilde{\chi}^0_1 \tilde{\chi}^0_3$\\ 
{\it Scenario~A1} (Fig.~4a):
Since $\tilde{\chi}^0_1$ is $\tilde{B}$--like and $\tilde{\chi}^0_3$ is 
higgsino--like  the cross sections are smaller than for 
$\tilde{\chi}^0_1 \tilde{\chi}^0_2$ production, reaching only 25~fb. Due to
the large $\tilde{\chi}^0_1 e \tilde{e}_R$ coupling one gets an 
enhancement by a factor of 2.7 for $P_{e^-}=+85\%$ and $P_{e^+}=-60\%$ 
compared with the unpolarized case.
 
{\it Scenario~B}: One gets a similar enhancement factor of about 2.8 as in 
scenario~A1. Due to slightly larger $\tilde{\chi}^0_3 e \tilde{e}_R$ coupling
the cross section is 42~fb for
$P_{e^-}=+85\%$ and $P_{e^+}=-60\%$.
\item $\tilde{\chi}^0_1 \tilde{\chi}^0_4$\\
One expects a behaviour similar to 
$e^+ e^-\to \tilde{\chi}^0_1 \tilde{\chi}^0_3$. In scenario~A1
the cross sections are about 
a factor 3 smaller and in scenario~B a factor 6, 
see Table~\ref{tab:2}.
\item $\tilde{\chi}^0_2 \tilde{\chi}^0_3$\\ 
{\it Scenario~A1} (see Fig.~4b):
Since the $\tilde{\chi}^0_2$ has a strong $\tilde{W}^3$ component 
one gets the largest cross 
sections with left polarized electrons and right polarized positrons.
With $P_{e^-}=-85\%$ and $P_{e^+}=+60\%$ the cross section is 
enhanced by a factor of 2.6 and reaches 41~fb.

{\it Scenario~B}: Due to slightly larger $\tilde{\chi}^0_3$ 
couplings than in scenario~A1 one gets a cross section of 79~fb for
$P_{e^-}=-85\%$ and $P_{e^+}=+60\%$.
\end{itemize}
Table~\re{tab:2} gives a survey of all cross sections 
$e^+e^-\to \tilde{\chi}^0_i \tilde{\chi}^0_j$ (including the invisible channel
$e^+ e^-\to \tilde{\chi}^0_1 \tilde{\chi}^0_1$) for different beam 
polarizations. Note the large cross sections for 
$e^+e^-\to \tilde{\chi}^0_3 \tilde{\chi}^0_4$ due to the large 
$Z^0$ couplings to the higgsino components.
In summary the cross sections 
can be enhanced by a factor two to three by polarizing both beams.
For pure gaugino--like neutralinos and
$m_{\tilde{e}_L}\gg m_{\tilde{e}_R}$ ($m_{\tilde{e}_L}\ll m_{\tilde{e}_R}$),
for  $P_{e^-}=+1$, $P_{e^+}=-1$ ($P_{e^-}=-1$, $P_{e^+}=+1$) the cross section
could even  be enlarged by a factor 4. For pure higgsino--like neutralinos
and $P_{e^-}=+1$, $P_{e^+}=-1$ ($P_{e^-}=-1$, $P_{e^+}=+1$) the enhancement 
factor is 1.7 (2.3) \ci{Gudi_diss,LC99_neut}.

If the polarizations of both beams are varied, the relative size of the cross 
sections strongly depends on the mixing character of both
neutralinos and on the selectron masses
\ci{Gudi_diss}. In particular, for 
$e^+ e^- \to \tilde{\chi}^0_i\tilde{\chi}^0_j$ and if
$\tilde{\chi}^0_i$ and $\tilde{\chi}^0_j$ are pure 
higgsinos, one obtains for $|P_{e^-}|=85\%$ and $|P_{e^+}|=60\%$ the sequence 
\bequ
\sigma^{-+}>\sigma^{+-}>\sigma^{-0}>
\sigma^{00}>\sigma^{+0}>\sigma^{--}>\sigma^{++}.
\la{eq_pol1}
\eequ

If $\tilde{\chi}^0_i$ and $\tilde{\chi}^0_j$ are 
pure gauginos, the order of the cross sections depends on the relative 
magnitude of the selectron masses $m_{\tilde{e}_L}$ and 
$m_{\tilde{e}_R}$.
For $m_{\tilde{e}_L}\gg m_{\tilde{e}_R}$ only right selectron exchange 
contributes, and one obtains
\bequ
\sigma^{+-}>\sigma^{+0}>\sigma^{00}>\sigma^{++}>\sigma^{--}>
\sigma^{-0}>\sigma^{-+},
\la{eq_pol3}
\eequ
whereas for $m_{\tilde{e}_R}\gg m_{\tilde{e}_L}$
(which may be realized in extended SUSY models,
\ci{HesselFranke} and references therein), one gets:  
\bequ
\sigma^{-+}>\sigma^{-0}>\sigma^{00}>\sigma^{--}>\sigma^{++}>
\sigma^{+0}>\sigma^{+-}.
\la{eq_pol2}
\eequ
A comparison of 
(\re{eq_pol1}) and (\re{eq_pol2}) shows that polarizing both beams
allows one to distinguish between a higgsino--like scenario and
a gaugino--like scenario with dominating $\tilde{e}_L$ exchange. 
This is not possible if only the electron beam is polarized. 

From Table~\re{tab:2}
one notices that for $e^+ e^-\to \tilde{\chi}^0_3 \tilde{\chi}^0_4$
one obtains the same ordering of the polarized cross
sections in scenario~A1 and B (with $\tilde{\chi}^0_3$ and
$\tilde{\chi}^0_4$ 
higgsino--like) as in (\re{eq_pol1}).
Comparing the sequence of cross sections for 
$e^+ e^-\to \tilde{\chi}^0_1 \tilde{\chi}^0_2$ in
scenario~A2 (with $\tilde{\chi}^0_1$, $\tilde{\chi}^0_2$ gaugino--like) 
with (\re{eq_pol3}), one observes the influence of $\tilde{e}_L$
exchange which is however suppressed due to the high mass 
$m_{\tilde{e}_L}=500$~GeV.
\subsection{Decay Lepton Forward--Backward Asymmetries\la{sec:33}}
We will discuss the forward--backward asymmetry of the
lepton angular distribution $d\sigma_e/d\cos\Theta_e$
in the overall c.m.s. of the combined reactions 
\bequ
e^+ e^-\to \tilde{\chi}^+_1 \tilde{\chi}^-_1,\quad\quad
\tilde{\chi}^-_1\to \tilde{\chi}^0_1 e^- \bar{\nu}
\la{eq_char}
\eequ
and 
\bequ
e^+ e^-\to \tilde{\chi}^0_1 \tilde{\chi}^0_2,\quad\quad
\tilde{\chi}^0_2\to \tilde{\chi}^0_1 e^+ e^-.
\la{eq_neut}
\eequ 
Here $\Theta_e$ denotes
the angle between the electron beam and the decay electron $e^-$.
The forward--backward asymmetry $A_{FB}$ of the decay electron is
defined as
\bequ
A_{FB}=\frac{\sigma_e(\cos\Theta_e>0)-\sigma_e(\cos\Theta_e<0)}
{\sigma_e(\cos\Theta_e>0)+\sigma_e(\cos\Theta_e<0)}, \la{eq_22}
\eequ
where 
\bequ
\sigma_e=\sigma(e^+ e^-\to \tilde{\chi}^+_1 \tilde{\chi}^-_1)\times
BR(\tilde{\chi}^-_1\to \tilde{\chi}^0_1 e^- \bar{\nu})
\la{eq_char2}
\eequ
for chargino production and decay and
\bequ
\sigma_e=\sigma(e^+ e^-\to \tilde{\chi}^0_1 \tilde{\chi}^0_2)\times
BR(\tilde{\chi}^0_2\to \tilde{\chi}^0_1 e^+ e^-)
\la{eq_neut2}
\eequ
for neutralino production and decay. Note that $d\sigma_e/d\cos\Theta_e$ and
$A_{FB}$ are sensitive to spin correlations.

The observable $A_{FB}$, eq.~(\re{eq_22}), 
is very sensitive to the gaugino component of the chargino/neutralino and 
the mass of the exchanged sneutrino or slepton.
It has the advantage of being independent of the parameters
of the squark sector which enter
in $\sigma_e$, eqs.~(\re{eq_char2}), (\re{eq_neut2}), 
via the leptonic branching ratio, which cancels in eq.~(\re{eq_22}). 
\subsubsection{$\sigma_e$ and 
$A_{FB}$ in Chargino Production and Decay \la{sec:331}}
Obviously, if the chargino $\tilde{\chi}^{\pm}_1$ has a substantial 
gaugino component, the sneutrino exchange in the t--channel has a strong 
influence on the cross section and angular distribution of chargino production.
In \cite{fpmt,Tsukamoto} the possibility of determining 
the sneutrino mass $m_{\tilde{\nu}_e}$ 
from the angular distribution of the production process $e^+e^- \to
\tilde{\chi}^+_1 \tilde{\chi}^-_1$ was studied. In the following
we study the
$m_{\tilde{\nu}_e}$ dependence of 
the decay lepton
forward--backward asymmetry $A_{FB}$, eq.~(\re{eq_22}),
in $e^+e^- \to \tilde{\chi}^+_1 \tilde{\chi}^-_1$, $\tilde{\chi}^-_1 \to
\tilde{\chi}^0_1 e^- \bar{\nu}_{e}$ \ci{Gudi_char,LC99_char}. 
As close to threshold this observable depends decisively 
on spin correlations, it is
instructive to have a closer look on its $m_{\tilde{\nu}_e}$
dependence. $A_{FB}$ also depends on the slepton mass 
$m_{\tilde{e}_L}$, due to the $\tilde{e}_L$ exchange in the
decay amplitude. Since $\tilde{\ell}_L$ and $\tilde{\nu}_{\ell}$ are
members of the same $SU(2)_L$ doublet, their masses are connected by
the relation
\cite{Tsukamoto,Ramond} 
\bequ
m^2_{\tilde{\ell}_L} = m^2_{\tilde{\nu}_{\ell}} - m^2_W \cos2\beta \la{eq_su2}
\eequ
with $m_W$ the mass of the $W^{\pm}$ boson. 
Relation (\re{eq_su2}) is fulfilled 
at tree level, and is only modified by
radiative corrections.

We first show in Fig.~5 the cross section $\sigma_e$,
eq.(\re{eq_char2}), as a function of $m_{\tilde{\nu}_e}$ at
$\sqrt{s}=2 m_{\tilde{\chi}^{\pm}_1}+10$~GeV, fixing $m_{\tilde{e}_L}$ by 
eq.~(\re{eq_su2}). $\sigma_e$ exhibits
a pronounced minimum, which is due to the destructive interference between 
$Z$ exchange and $\tilde{\nu}_e$ exchange \ci{Gudi_char}.
For $\sqrt{s}$ near threshold this minimum is
approximately at $m_{\tilde{\nu}_e} \approx m_{\tilde{\chi}^{\pm}_1}$
and in the limit $\sqrt{s} \to 2m_{\tilde{\chi}^{\pm}_1}$ 
the minimum reaches exactly $m_{\tilde{\nu}_e} \to m_{\tilde{\chi}^{\pm}_1}$. 
Due to this minimum there is an ambiguity when one tries to
determine $m_{\tilde{\nu}_e}$
by measuring $\sigma_e$. At higher $\sqrt{s}$ the minimum is shifted to 
higher values of $m_{\tilde{\nu}}$ \ci{Gudi_diss,LC99_char}.

In Fig.~\re{fig_5}a and b we show the contour lines of $A_{FB}$ as a function
of $m_{\tilde{\nu}}$ and $m_{\tilde{e}_L}$
at $\sqrt{s}=2 m_{\tilde{\chi}^{\pm}_1}+10$~GeV and $\sqrt{s}=500$~GeV,
respectively, for $P_{e^-}=-85$\% and $P_{e^+}=+60\%$. In order to study
separately the $m_{\tilde{\nu}_e}$ and $m_{\tilde{e}_L}$ dependence of $A_{FB}$
we have relaxed the mass relation eq.(\re{eq_su2}).
Apart from $m_{\tilde{\nu}}$ and $m_{\tilde{e}_L}$ the parameters are as in 
scenario~A1. The large asymmetry is mainly due to $\tilde{\nu}_e$ exchange
in the crossed channel of the production.
Close to threshold also the $\tilde{e}_L$ exchange in the 
decay $\tilde{\chi}^-_1\to \tilde{\chi}^0_1 e^- \bar{\nu}_e$
plays an important role, see Fig.~\re{fig_5}a.
At energies far from threshold the charginos 
have a large energy, 
and the decay lepton has essentially the same 
direction as the chargino \ci{Feng}. Therefore,  
the $\tilde{e}_L$ dependence is weaker
\ci{LC99_char}, see Fig.~\re{fig_5}b.

Due to the dominance of the t--channel contribution 
in scenario~A1 the polarization dependence of the nominator 
and denominator in the ratio, eq.~(\re{eq_22}), almost cancels. 
Therefore $A_{FB}$ depends only weakly on the polarizations
of the beams. 
 
Turning now to the question how accurate the sneutrino mass
$m_{\tilde{\nu}_e}$ can be determined from chargino pair production
and decay, Figs.~\re{fig_5}a and b show that there is an appreciable 
$m_{\tilde{\nu}_e}$ dependence of $A_{FB}$.
We first consider the case $m_{\tilde{\nu}_e} \grts
\sqrt{s}/2$, where $\tilde{\nu}_e \bar{\tilde{\nu}}_e$ pair
production is kinematically not possible. 
We assume that the slepton mass $m_{\tilde{e}_L}$ and the
other SUSY parameters are known with good precision.
For definiteness, we take, $m_{\tilde{e}_L}=200$~GeV, and the
other SUSY parameters as in scenario A1. 

At $\sqrt{s}=500$~GeV, with a luminosity of
${\cal L}=500$~fb$^{-1}$, 
$A_{FB}$ can be measured up to $<\pm 1\%$, if we take 
only the statistical error $\delta(A_{FB})$. 
This means that in the range $350$~GeV $\lets m_{\tilde{\nu}_e}
\lets 800$~GeV an accuracy of about $|\delta m_{\tilde{\nu}_e}|< 10$~GeV 
may be achieved. The experimental errors 
of $m_{\tilde{e}_L}$ and the other SUSY parameters are neglected.
The ambiguity at $\sqrt{s}=500$~GeV (Fig.~\re{fig_5}b) in the range
$250$~GeV $\lets m_{\tilde{\nu}_e} \lets 350$~GeV 
can be resolved by measuring $A_{FB}$ at
different c.m.s. energies. Similarly, at
$\sqrt{s} = 2m_{\tilde{\chi}^{\pm}_1}+10$~GeV (Fig.~\re{fig_5}a), 
$A_{FB}$ is quite 
sensitive to $m_{\tilde{\nu}_e}$ in the range $135$~GeV$\lets
m_{\tilde{\nu}_e} \lets 350$~GeV, where direct production is again 
not possible. 
If only the electron beam is polarized,  
$\delta(A_{FB})$ would be larger by about 20\% in the case considered.
Moreover, for $P_{e^-}=-85\%$ and $P_{e^+}=+60\%$
we obtain $P_{eff}=96\%$. If $P_{e^-}$ and $P_{e^+}$ have an error
of $\pm 1\%$, the error of 
$P_{eff}$ would only be $\pm 0.83\%$.
For a more quantitative assessment of the accuracy of 
$m_{\tilde{\nu}_e}$ that can be expected from measuring the decay
lepton forward--backward asymmetry in chargino production, Monte
Carlo studies taking into account experimental cuts and detector
simulation would be necessary. For instance a cut  
$-0.9<\cos\Theta_{e}<0.9$ would lead to about 10\% smaller values of 
$\sigma_e$ and  $A_{FB}$.

In case $m_{\tilde{\nu}_{e}} < \sqrt{s}/2$,
$\tilde{\nu}_{e} \bar{\tilde{\nu}}_{e}$ pairs can be directly
produced. If 
$m_{\tilde{\chi}^{\pm}_1} < m_{\tilde{\nu}_{e}} < \sqrt{s}/2$, then the 
visible decay $\tilde{\nu}_{e} \to e^- \tilde{\chi}^+_1$ is
kinematically allowed, and will presumably have a sufficiently
high branching ratio. We do not treat this case here, because
measuring the cross section of  
$e^+e^- \to \tilde{\nu}_{e} \bar{\tilde{\nu}}_{e}$ at threshold will
allow us to determine $m_{\tilde{\nu}_{e}}$ with good
accuracy \cite{martyn}. If 
$m_{\tilde{\nu}_{e}} < m_{\tilde{\chi}^{\pm}_1} < \sqrt{s}/2$, then 
$\tilde{\nu}_{e}$ has no visible decay with sufficiently high
branching ratio. However, the two--body chargino decay 
$\tilde{\chi}^{\pm}_1 \to e^{\pm} \stackrel{(-)}{\tilde{\nu}_{e}}$ is possible.
Measuring the endpoints of the energy spectrum of the decay
leptons $e^+$ and $e^-$ will provide a very precise
determination of 
the masses $m_{\tilde{\chi}^{\pm}_1}$ and $m_{\tilde{\nu}_{e}}$. 
The alternative method 
to determine $m_{\tilde{\nu}_{e}}$ by measuring the decay lepton 
forward--backward asymmetry $A_{FB}$ of chargino production
will, in principle, also be possible. However, the accuracy of
$m_{\tilde{\nu}_{e}}$ obtainable in this way is expected to be
lower than that from the decay lepton energy spectrum.
\subsubsection{$A_{FB}$ in Neutralino Production and Decay}\la{sec:332}
Owing to the Majorana character of the neutralinos the angular distribution 
of the production process is symmetric under the exchange
$\cos\Theta\leftrightarrow -\cos\Theta$, where $\Theta$ is the production 
angle of $\tilde{\chi}^0_j$ \ci{Christova}.
The angular distribution of the decay lepton $\ell^-$, 
however, depends on the polarization of $\tilde{\chi}^0_j$.
Since the longitudinal polarization $\Sigma^3_P$ and the transverse 
polarization $\Sigma^1_P$ of $\tilde{\chi}^0_j$ are forward--backward 
antisymmetric, the lepton forward--backward asymmetry $A_{FB}$ 
of the decay lepton, eq.~(\re{eq_22}), may become quite large.
We will plot $A_{FB}$ not too far from threshold because  it 
decreases with $\sqrt{s}$ for fixed neutralino masses.

In Figs.~7a we show $A_{FB}$ of 
the decay electron for reaction (\re{eq_neut})
as a function of the beam polarizations
 for scenario~A1 at
$\sqrt{s}=(m_{\tilde{\chi}^0_1}+m_{\tilde{\chi}^0_2})+30$~GeV. 
As one can see, by appropriately polarizing both beams one gets
a larger asymmetry. Note that in this scenario 
$A_{FB}$ is practically zero if both beams are unpolarized.

In Figs.~7b and c we show the contour lines of $A_{FB}$ as
a function of $m_{\tilde{e}_L}$ and $m_{\tilde{e}_R}$ at
$\sqrt{s}=m_{\tilde{\chi}^0_1}+m_{\tilde{\chi}^0_2}+30$~GeV,
for $P_{e^-}=-85\%$, $P_{e^+}=+60\%$ and $P_{e^-}=+85\%$, 
$P_{e^+}=-60\%$, respectively. The other parameters are as in scenario~A1. For 
$m_{\tilde{e}_L}\approx m_{\tilde{e}_R}$ the asymmetry $A_{FB}$ is very small 
for both polarizations considered, but can reach approximately 
$\pm 23\%$ if $m_{\tilde{e}_L}\neq m_{\tilde{e}_R}$.
Measuring the lepton forward--backward asymmetry $A_{FB}$ in addition to the 
total cross section will give constraints 
on the selectron masses $m_{\tilde{e}_L}$
and $m_{\tilde{e}_R}$. 

In the following we will estimate the precision to be expected if one tries 
to constrain the selectron masses by the data of $A_{FB}$ of reaction (22).
We assume that the right selectron mass
is known with good precision and, for definiteness, we take
$m_{\tilde{e}_R}=240$~GeV. We further assume that the measurement of $A_{FB}$
has given $A_{FB}=-10$\% at 
$\sqrt{s}=m_{\tilde{\chi}^0_1}+m_{\tilde{\chi}^0_2}+30$~GeV. 
For an integrated luminosity ${\cal L}=500$~fb$^{-1}$ the statistical error is
expected to be 
$\delta(A_{FB})\approx \pm 3$\%. Fig.~7b shows that, for
$P_{e^-}=-85\%$ and $P_{e^+}=+60\%$, $m_{\tilde{e}_L}$ has to be either
in the range $280$~GeV$<m_{\tilde{e}_L}<320$~GeV or in  
$410$~GeV$<m_{\tilde{e}_L}<460$~GeV. This example demonstrates that the
mass of $\tilde{e}_L$ can be constrained by measuring $A_{FB}$ even if it is
too heavy to be directly produced in $e^+ e^-$ annihilation.
As a second example we assume that $m_{\tilde{e}_R}$ is known to be
$m_{\tilde{e}_R}=340$~GeV, and $A_{FB}=+15\%$
has been measured. This constrains $m_{\tilde{e}_L}$ 
to be either in the mass region
250~GeV$<m_{\tilde{e}_L}<300$~GeV or in
$450$~GeV$<m_{\tilde{e}_L}<600$~GeV.

As in the chargino case, if both beams are polarized 
the cross sections can be enhanced 
and therefore the statistical error
of $A_{FB}$ can be reduced. Moreover,
also the error of the effective polarization $\delta(P_{eff})$ can be reduced.

In \ci{LC99_neut} we also studied the $\sqrt{s}$
dependence of the lepton forward--backward asymmetry 
$A_{FB}$. For $\sqrt{s}\gg (m_{\tilde{\chi}^0_1}+m_{\tilde{\chi}^0_2})$ the
angular distribution of the decay lepton
is essentially the same as that of the  
neutralino $\tilde{\chi}^0_2$ \ci{Feng}.
Due to the Majorana character of the
decaying neutralino $\tilde{\chi}^0_2$
the lepton forward--backward asymmetry now practically vanishes.
\subsubsection{$M_1$ Dependence of $\sigma_e$, $A_{FB}$ and $A_{pol}$
in Neutralino Production and Decay}\la{sec:333}
As is well known, the neutralino masses as well as the 
$Z^0 \tilde{\chi}^0_i \tilde{\chi}^0_j$ couplings 
and the $\tilde{\chi}^0_i \tilde{\ell} \ell$
couplings also depend on the gaugino mass parameter $M_1$ 
\ci{Gudi_neut,LC99_neut}.
So far we have used the GUT relation (\re{eq_m1})
for the gaugino masses.
In the following we will be more general and not use
this relation \ci{Choi,Kneur,Feng,Snowmass,Claus}. We will discuss the 
$M_1$ dependence of the cross section, the polarization asymmetry 
and the forward--backward asymmetry
of the decay electron \ci{Gudi_neut, LC99_neut} in the reaction
$e^+ e^-\to \tilde{\chi}^0_1 \tilde{\chi}^0_2$, 
$\tilde{\chi}^0_2\to \tilde{\chi}^0_1 e^+ e^-$.
All other parameters are chosen as in scenario~A1 except the mass 
of $\tilde{e}_R$, which we take $m_{\tilde{e}_R}=161$~GeV.

\bsl
Fig.~\re{fig_8}a exhibits the $M_1$ dependence of $\sigma(e^+e^-\to 
\tilde{\chi}^0_1\tilde{\chi}^0_2)\times BR(\tilde{\chi}^0_2\to e^+ e^- 
\tilde{\chi}^0_1)$ at $\sqrt{s}=(m_{\tilde{\chi}^0_1}+m_{\tilde{\chi}^0_2})+
30$~GeV for $m_{\tilde{e}_R}=161$~GeV and $m_{\tilde{e}_L}=176$~GeV
in the region 40 GeV$<M_1<m_{\tilde{e}_R}$ 
for various beam polarizations \ci{LC99_neut}.
We do not consider values $|M_1|>m_{\tilde{e}_R}$, where 
$m_{\tilde{\chi}^0_2}>m_{\tilde{\ell}_R}$, because then the two--body decay
$\tilde{\chi}^0_2\to \tilde{\ell}_R+\ell$ would be the dominant decay 
channel. 
Fig.~\re{fig_8}b shows the analogous curves for $m_{\tilde{e}_L}=500$~GeV. 
One clearly sees from the curves for left polarized electrons and/or
right polarized positrons that the 
$\tilde{e}_L$ exchange is strongly suppressed, and one obtains higher cross 
sections for right polarized $e^-$ beams.
\esl

We have also studied the $M_1$ 
dependence of the forward--backward 
asymmetry $A_{FB}$ of the decay electron, 
eq.~(\re{eq_22}), shown in Fig.~\re{fig_9}a and b for 
$m_{\tilde{e}_L}=176$~GeV and $m_{\tilde{e}_L}=500$~GeV. 
One notices a significant 
variation with $M_1$ and a strong dependence on the beam 
polarizations. Comparing Fig.~\re{fig_9}a and b,
one observes that in the region 40~GeV$<M_1<$100~GeV
the forward--backward asymmetries 
of the decay electron are different. This is 
due to the suppression of $\tilde{e}_L$ exchange in 
Fig.~\re{fig_9}b. 

Another observable which has a characteristic $M_1$ dependence is the 
polarization asymmetry defined as 
\bequ
A_{pol}=\frac{\sigma^{AB}-\sigma^{CD}}
{\sigma^{AB}+\sigma^{CD}}=
\frac{\sigma_e^{AB}-\sigma_e^{CD}}
{\sigma_e^{AB}+\sigma_e^{CD}}\quad,\la{A_pol}
\eequ
where $A,C$ indicate two values of 
the polarization $P_{e^-}$, and $B,D$ two values of the polarization 
$P_{e^+}$. 

In Fig.~10 we show $A_{pol}$ as a function of $M_1$ at 
$\sqrt{s}=m_{\tilde{\chi}^0_1}+m_{\tilde{\chi}^0_2}+30$~GeV for different
beam polarizations. Apart from $M_1$ we take the parameters and 
masses as in scenario~A1. 
In the case of a small mass difference between $\tilde{e}_L$ and $\tilde{e}_R$,
$M_1$ can be constrained by measuring $A_{pol}$ for different beam 
polarizations. For larger selectron mass differences the $M_1$ 
dependence of $A_{pol}$ is much weaker. 
\section{Conclusions}
The objective of this paper has been twofold. Firstly, we have studied 
the advantage of having both the $e^-$ and the $e^+$ beam polarized. 
If the polarizations of $e^-$ and $e^+$ are varied,
the cross sections depend significantly 
on the mixing character of the charginos and neutralinos and on 
the masses of $\tilde{\nu}_e$, $\tilde{e}_L$ and $\tilde{e}_R$.
By an appropriate choice of polarizations
one can obtain up to three times larger cross sections than in the unpolarized
case. 
 Secondly, we have studied the forward--backward asymmetry of 
the decay electron in
$e^+ e^-\to \tilde{\chi}^+_1 \tilde{\chi}^-_1$, 
$\tilde{\chi}^-_1\to \tilde{\chi}^0_1 e^- \bar{\nu}_e$, and in
$e^+ e^-\to \tilde{\chi}^0_1 \tilde{\chi}^0_2$, 
$\tilde{\chi}^0_2\to \tilde{\chi}^0_1 e^+ e^-$
taking into account the full spin correlations between production 
and decay. 
Measuring this asymmetry 
for various beam polarizations gives further constraints on the masses of 
$\tilde{\nu}_e$, 
$\tilde{e}_L$ and $\tilde{e}_R$, 
also if direct production of these particles is kinematically not possible.
It also constrains the mixing properties of 
the charginos and neutralinos. 
We have also studied the dependence on the gaugino mass parameter $M_1$. For a 
determination of $M_1$ the use of polarized $e^+$ and 
$e^-$ beams would also be very useful. 
\vspace{-1.2mm}
\section*{Acknowledgments}
\vspace{-3mm}
\bsl
We thank U.~Martyn for many valuable discussions.
Parts of the calculations have been performed on the QCM cluster at the
University of Karlsruhe, supported by the Deutsche Forschungsgemeinschaft
under contract number FOR 264/2-1. We are grateful to W.~Porod and 
S.~Hesselbach 
for providing the computer programs for neutralino and chargino total widths.
This work was also supported by 
the German Federal Ministry for
Research and Technology (BMBF) under contract number
05 7WZ91P (0), by the Deutsche Forschungsgemeinschaft under
contract Fr 1064/4-1, and the `Fonds zur
F\"orderung der wissenschaftlichen Forschung' of Austria, Project
\nopagebreak No. P13139-PHY. 
\esl\nopagebreak

\begin{table}[t]
\begin{center}
{\small 
\begin{tabular}{|l||c|c|c|c|c|c|c|c|c|c|c|c|}
 & $M_2$ & $\mu$ & $\tan\beta$ &
$m_{\tilde{e}_L}$ & $m_{\tilde{e}_R}$ & $m_{\tilde{\nu}}$ &
$m_{\tilde{\chi}^{0}_1}$ &
$m_{\tilde{\chi}^{0}_2}$ & $m_{\tilde{\chi}^0_3}$ &
$m_{\tilde{\chi}^0_4}$ & $m_{\tilde{\chi}^{\pm}_1}$ &
$m_{\tilde{\chi}^{\pm}_2}$  \\ \hline
 A1 & 152 & 316 & 3 & 176 & 132 & 161 & 71 & 130 & 320 & 348 & 128 & 346 \\ \hline
 A2 & 152 & 316 & 3 & 500 & 132 & 161 & 71 & 130 & 320 & 348 & 128 & 346\\ \hline 
 B & 150 & 263 & 30 & 217 & 183 & 202 & 75 & 133 & 273 & 293 & 132 & 295  \\ \hline
\end{tabular}
\caption{Parameters and masses (in GeV)
in scenarios A1, A2, and B. \label{tab:1}}
}
\end{center}
\begin{center}
{\small
\begin{tabular}{|lc||c|c|c|c|c|c|c|}
\hline
A1& $\tilde{\chi}^+_1\tilde{\chi}^-_1${\footnotesize ($\sqrt{s}=266$ GeV)}
& $(-+)$ & $(-0)$ &$(00)$ &$(--)$ & $(++)$ &$(+0)$ & $(+-)$\\ 
 & $\sigma$/fb & 410 & 256 & 139 & 103  & 34  & 22 & 10 \\ \hline
& $\tilde{\chi}^+_1\tilde{\chi}^-_2${\footnotesize ($\sqrt{s}=484$ GeV)} & $(+-)$ & $(+0)$ &$(00)$ &$(-+)$ & $(-0)$ &$(++)$ & $(--)$\\ 
& $\sigma$/fb & 6.0  & 3.8 & 2.9 & 2.7 & 1.9  & 1.7 & 1.1 \\ \hline
& $\tilde{\chi}^+_2\tilde{\chi}^-_2$ {\footnotesize ($\sqrt{s}=702$ GeV)} & $(-+)$ & $(-0)$ &$(00)$ &$(--)$ & $(+-)$ &$(+0)$ & $(++)$\\ 
& $\sigma$/fb & 113  & 72 & 45 & 30 & 24  & 19 & 15 \\ \hline\hline
B& $\tilde{\chi}^+_1\tilde{\chi}^-_1$ {\footnotesize ($\sqrt{s}=274$ GeV)} & $(-+)$ & $(-0)$ &$(00)$ &$(--)$ & $(++)$ &$(+0)$ & $(+-)$\\ 
& $\sigma$/fb &709 & 443 & 239 & 177  & 57  & 36 & 15 \\ \hline
& $\tilde{\chi}^+_1\tilde{\chi}^-_2$ {\footnotesize ($\sqrt{s}=437$ GeV)}  &$(+-)$ & $(+0)$ &$(00)$ &$(++)$ & $(-+)$ &$(-0)$ & $(--)$\\ 
& $\sigma$/fb &10.0  & 6.3 & 3.9 & 2.6 & 1.9  & 1.6 & 1.2 \\ \hline
& $\tilde{\chi}^+_2\tilde{\chi}^-_2$ {\footnotesize ($\sqrt{s}=600$ GeV)} & $(-+)$ & $(-0)$ &$(00)$ &$(--)$ & $(+-)$ &$(+0)$ & $(++)$\\ 
& $\sigma$/fb & 194  & 123 & 75 & 51 & 32  & 28 & 23 \\ \hline
\end{tabular}
\caption{{Polarized cross sections 
$\sigma=\sigma(e^+ e^-\to \tilde{\chi}^+_i \tilde{\chi}^-_j)$/fb, i, j=1, 2, 
at $\sqrt{s}=m_{\tilde{\chi}^{+}_i}+m_{\tilde{\chi}^-_j}+10$~GeV
in scenarios~A1 and B, see Table~\re{tab:1}, for unpolarized beams 
$(00)$, only electron beam polarized $(-0)$, $(+0)$ with  $P_{e^-}=\pm 85\%$
and both beams polarized with 
$P_{e^-}=\pm 85\%$, $P_{e^+}=\pm 60\%$. }\la{table_11}}
}
\end{center}
\end{table}
\begin{table}
{\small\vspace{-.5cm}
\begin{tabular}{|lc||c|c|c|c|c|c|c|}
\hline
A1& $\tilde{\chi}^0_1 \tilde{\chi}^0_1${\footnotesize ($\sqrt{s}=172$ GeV)} & $(+-)$ & $(+0)$ &$(00)$ &$(++)$ & $(--)$ &$(-0)$ & $(-+)$\\ 
& $\sigma$/fb & 430 & 269 & 146 & 108  & 35  & 23 & 11 \\ \hline
& $\tilde{\chi}^0_1 \tilde{\chi}^0_2${\footnotesize ($\sqrt{s}=231$ GeV)} & $(-+)$ & $(-0)$ &$(+-)$ &$(00)$ & $(+0)$ &$(--)$ & $(++)$\\ 
& $\sigma$/fb & 56  & 36 & 35 & 30 & 25  & 17 & 13 \\ \hline
 & $\tilde{\chi}^0_1 \tilde{\chi}^0_3$ {\footnotesize ($\sqrt{s}=421$ GeV)}& $(+-)$ & $(+0)$ &$(00)$ &$(++)$ & $(-+)$ &$(-0)$ & $(--)$\\ 
& $\sigma$/fb & 25  & 16 & 9.3 & 6.5 & 3.0  & 2.8 & 2.7 \\ \hline
& $\tilde{\chi}^0_1 \tilde{\chi}^0_4$ {\footnotesize ($\sqrt{s}=449$ GeV)}& $(+-)$ & $(+0)$ &$(00)$ &$(++)$ & $(-+)$ &$(-0)$ & $(--)$\\ 
& $\sigma$/fb & 7.9  & 5.0 & 3.2 & 2.1 & 1.7  & 1.4 & 1.0 \\ \hline
& $\tilde{\chi}^0_2 \tilde{\chi}^0_2$ {\footnotesize ($\sqrt{s}=290$ GeV)}& $(-+)$ & $(-0)$ &$(00)$ &$(--)$ & $(++)$ &$(+0)$ & $(+-)$\\ 
& $\sigma$/fb & 121  & 76 & 41 & 30 & 10  & 6.3 & 2.7 \\ \hline
& $\tilde{\chi}^0_2 \tilde{\chi}^0_3$ {\footnotesize ($\sqrt{s}=480$ GeV)}& $(-+)$ & $(-0)$ &$(00)$ &$(--)$ & $(+-)$ &$(+0)$ & $(++)$\\ 
& $\sigma$/fb & 41  & 26 & 16 & 11 & 7.2  & 6.1 & 4.9 \\ \hline
& $\tilde{\chi}^0_2 \tilde{\chi}^0_4$ {\footnotesize ($\sqrt{s}=508$ GeV)}& $(-+)$ & $(-0)$ &$(00)$ &$(--)$ & $(++)$ &$(+0)$ & $(+-)$\\ 
& $\sigma$/fb & 11  & 6.6 & 3.6 & 2.7 & 0.9  & 0.6 & 0.4 \\ \hline
& $\tilde{\chi}^0_3 \tilde{\chi}^0_3$ {\footnotesize ($\sqrt{s}=670$ GeV)}& $(-+)$ & $(+-)$ &$(-0)$ &$(00)$ & $(+0)$ &$(--)$ & $(++)$\\ 
& $\sigma$/fb & \multicolumn{7}{|c|}{$<10^{-2}$~fb}\\ \hline
& $\tilde{\chi}^0_3 \tilde{\chi}^0_4$ {\footnotesize ($\sqrt{s}=696$ GeV)}& $(-+)$ & $(+-)$ &$(-0)$ &$(00)$ & $(+0)$ &$(--)$ & $(++)$\\ 
& $\sigma$/fb & 60  & 41 & 39 & 34 & 28  & 18 & 15 \\ \hline
& $\tilde{\chi}^0_4 \tilde{\chi}^0_4$ {\footnotesize ($\sqrt{s}=722$ GeV)}& $(-+)$ & $(-0)$ &$(00)$ &$(--)$ & $(+-)$ &$(+0)$ & $(++)$\\ 
& $\sigma$/fb & \multicolumn{7}{|c|}{$<0.2$~fb} \\ \hline\hline
A2 & $\tilde{\chi}^0_1 \tilde{\chi}^0_2$ {\footnotesize ($\sqrt{s}=231$ GeV)}&$ (+-)$ & $(+0)$ &$(00)$ &$(++)$ & $(-+)$ &$(-0)$ & $(--)$\\ 
& $\sigma$/fb & 35  & 22 & 13 & 9.0 & 4.7  & 4.2 & 3.8 \\ \hline\hline
B&$\tilde{\chi}^0_1 \tilde{\chi}^0_1$ {\footnotesize ($\sqrt{s}=180$ GeV)} & $(+-)$ & $(+0)$ &$(00)$ &$(++)$ & $(--)$ &$(-0)$ & $(-+)$\\ 
& $\sigma$/fb & 191 & 119 & 65 & 48  & 16  & 10 & 5.7 \\ \hline
& $\tilde{\chi}^0_1 \tilde{\chi}^0_2$ {\footnotesize ($\sqrt{s}=238$ GeV)}& $(-+)$ & $(-0)$ &$(00)$ &$(+-)$ & $(--)$ &$(+0)$ & $(++)$\\ 
& $\sigma$/fb & 50  & 31 & 21 & 14 & 13  & 11 & 7 \\ \hline
& $\tilde{\chi}^0_1 \tilde{\chi}^0_3$ {\footnotesize ($\sqrt{s}=378$ GeV)}& $(+-)$ & $(+0)$ &$(00)$ &$(++)$ & $(--)$ &$(-0)$ & $(-+)$\\ 
& $\sigma$/fb & 42  & 26 & 15 & 11 & 3.9  & 3.4 & 2.9 \\ \hline
& $\tilde{\chi}^0_1 \tilde{\chi}^0_4$ {\footnotesize ($\sqrt{s}=398$ GeV)}& $(+-)$ & $(+0)$ &$(00)$ &$(-+)$ & $(++)$ &$(-0)$ & $(--)$\\ 
& $\sigma$/fb & 6.3  & 4.0 & 2.7 & 1.8 & 1.7  & 1.4 & 1.0 \\ \hline
& $\tilde{\chi}^0_2 \tilde{\chi}^0_2$ {\footnotesize ($\sqrt{s}=296$ GeV)}& $(-+)$ & $(-0)$ &$(00)$ &$(--)$ & $(++)$ &$(+0)$ & $(+-)$\\ 
& $\sigma$/fb & 59  & 37 & 20 & 15 & 4.8  & 3.1 & 1.3 \\ \hline
& $\tilde{\chi}^0_2 \tilde{\chi}^0_3$ {\footnotesize ($\sqrt{s}=436$ GeV)}& $(-+)$ & $(-0)$ &$(00)$ &$(--)$ & $(+-)$ &$(+0)$ & $(++)$\\ 
& $\sigma$/fb & 79  & 50 & 31 & 21 & 16  & 13 & 10 \\ \hline
& $\tilde{\chi}^0_2 \tilde{\chi}^0_4$ {\footnotesize ($\sqrt{s}=456$ GeV)}& $(-+)$ & $(-0)$ &$(00)$ &$(--)$ & $(++)$ &$(+0)$ & $(+-)$\\ 
& $\sigma$/fb & 11  & 6.9 & 3.7 & 2.7 & 1.0  & 0.6 & 0.3 \\ \hline
& $\tilde{\chi}^0_3 \tilde{\chi}^0_3$ {\footnotesize ($\sqrt{s}=576$ GeV)}& $(-+)$ & $(+-)$ &$(-0)$ &$(00)$ & $(+0)$ &$(--)$ & $(++)$\\ 
& $\sigma$/fb & \multicolumn{7}{|c|}{$<10^{-2}$~fb} \\ 
\hline
& $\tilde{\chi}^0_3 \tilde{\chi}^0_4$ {\footnotesize ($\sqrt{s}=596$ GeV)}& $(-+)$ & $(+-)$ &$(-0)$ &$(00)$ & $(+0)$ &$(--)$ & $(++)$\\ 
& $\sigma$/fb & 78  & 55 & 51 & 44 & 37  & 24 & 20 \\ \hline
& $\tilde{\chi}^0_4 \tilde{\chi}^0_4$ {\footnotesize ($\sqrt{s}=616$ GeV)}& $(-+)$ & $(-0)$ &$(00)$ &$(--)$ & 
$(+-)$ &$(+0)$ & $(++)$\\ 
& $\sigma$/fb & \multicolumn{7}{|c|}{$<0.4$~fb}
\\ \hline
\end{tabular} }\vspace{-.4cm}
\caption{Polarized cross sections 
$\sigma=\sigma(e^+ e^-\to \tilde{\chi}^0_i \tilde{\chi}^0_j)$/fb,
i, j=1,$\ldots$,4, 
at $\sqrt{s}= m_{\tilde{\chi}^{0}_i}+m_{\tilde{\chi}^0_j}+30$~GeV
in scenarios~A1,~A2 and B, see Table~\re{tab:1}, for unpolarized beams 
$(00)$, only electron beam polarized $(-0)$, $(+0)$ with  $P_{e^-}=\pm 85\%$
and both beams polarized with 
$P_{e^-}=\pm 85\%$, $P_{e^+}=\pm 60\%$. \la{tab:2}}
\end{table}
\begin{figure}[t]
\begin{minipage}{3.5cm}
\begin{center}
{\setlength{\unitlength}{1cm}
\hspace*{.7cm}
\begin{picture}(3.5,2.5)
\put(0,-1.1){\includegraphics{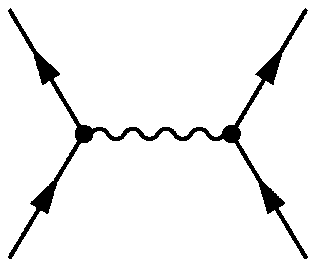}}
\put(-.7,-1){$e^{-}$}
\put(1.7,-1){$\tilde{\chi}^{-}_j$}
\put(-.7,1.7){$e^{+}$}
\put(1.7,1.7){$\tilde{\chi}^{+}_i$}
\put(.5,.7){$\gamma$}
\end{picture}}
\end{center}
\end{minipage}
\hspace{1.5cm}
\vspace{.4cm}
\begin{minipage}{3cm}
\begin{center}
{\setlength{\unitlength}{1cm}
\begin{picture}(3,2.5)
\put(0,-1.1){\includegraphics{prog.ps}}
\put(-.7,-1){$e^{-}$}
\put(1.7,-1){$\tilde{\chi}^{-}_j$}
\put(-.7,1.7){$e^{+}$}
\put(1.7,1.7){$\tilde{\chi}^{+}_i$}
\put(.4,.7){$Z^0$}
\end{picture}}
\end{center}
\end{minipage}
\hspace{1.3cm}
\vspace{.8cm}
\begin{minipage}{3cm}
\begin{center}
{\setlength{\unitlength}{1cm}
\begin{picture}(3,2)
\put(0,-1.4){\includegraphics{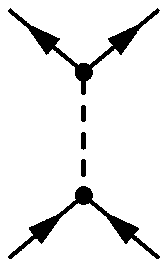}}
\put(-.7,-1.2){$e^{-}$}
\put(-.7,1.4){$e^{+}$}
\put(1.1,-1.2){$\tilde{\chi}^{-}_j$}
\put(1.1,1.4){$\tilde{\chi}^{+}_i$}
\put(-.2,.2){$\tilde{\nu}_e$}
\end{picture}}
\end{center}
\end{minipage}
\begin{minipage}{3.5cm}
\begin{center}
{\setlength{\unitlength}{1cm}
\hspace*{.7cm}
\begin{picture}(3.5,2.5)
\put(0,-1.1){\includegraphics{prog.ps}}
\put(-.7,-1){$e^{-}$}
\put(1.7,-1){$\tilde{\chi}^{0}_j$}
\put(-.7,1.7){$e^{+}$}
\put(1.7,1.7){$\tilde{\chi}^{0}_i$}
\put(.5,.7){$Z^0$}
\end{picture}}
\end{center}
\end{minipage}
\hspace{-.4cm}
\vspace{.8cm}
\begin{minipage}{3cm}
\begin{center}
{\setlength{\unitlength}{1cm}
\begin{picture}(3,2)
\put(0,-1.5){\includegraphics{prosn.ps}}
\put(-.7,-1.2){$e^{-}$}
\put(-.7,1.4){$e^{+}$}
\put(1.1,-1.2){$\tilde{\chi}^{0}_j$}
\put(1.1,1.4){$\tilde{\chi}^{0}_i$}
\put(-.7,.2){$\tilde{e}_{L,R}$}
\end{picture}}
\end{center}
\end{minipage}
\begin{minipage}{3cm}
\begin{center}
{\setlength{\unitlength}{1cm}
\begin{picture}(3,2)
\put(0,-1.5){\includegraphics{prosn.ps}}
\put(-.7,-1.2){$e^{-}$}
\put(-.7,1.4){$e^{+}$}
\put(1.1,-1.2){$\tilde{\chi}^{0}_i$}
\put(1.1,1.4){$\tilde{\chi}^{0}_j$}
\put(-.7,.2){$\tilde{e}_{L,R}$}
\end{picture}}
\end{center}
\end{minipage}
\vspace{.8cm}
\begin{picture}(10,1)
\put(0,-.1){Figure 1a: Feynman diagrams for chargino production 
$e^+ e^- \to \tilde{\chi}^+_i \tilde{\chi}^-_j$ and} 
\put(.1,-.6){neutralino production 
$e^+ e^- \to \tilde{\chi}^0_i \tilde{\chi}^0_j$.}
\end{picture}
\begin{minipage}[h]{3cm}
\begin{center}
{\setlength{\unitlength}{1cm}
\begin{picture}(3,2.5)
\put(.8,-.1){\includegraphics{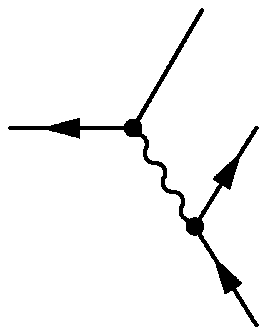}}
\put(2.6,-1){$\stackrel{(-)}{\nu_{\ell}}$}
\put(.2,1.2){$\tilde{\chi}^{\pm}_{i}$}
\put(2.6,.9){$\ell^{\pm}$}
\put(2.2,1.9){$\tilde{\chi}^0_{k}$}
\put(.8,.1){$W^{\pm}$}
\end{picture}}
\end{center}
\end{minipage}
\hspace{1cm}
\vspace{.8cm}
\begin{minipage}[h]{3cm}
\begin{center}
{\setlength{\unitlength}{1cm}
\begin{picture}(3,2.9)
\put(+.3,+0){\includegraphics{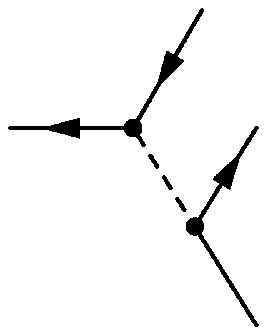}}
\put(.5,2.1){$\stackrel{(-)}{\nu_{\ell}}$}
\put(.9,1.1){$\ell^{\pm}$}
\put(-1.5,1.4){$\tilde{\chi}^{\pm}_{i}$}
\put(.9,-.8){$\tilde{\chi}^0_{k}$}
\put(-.6,.5){$\tilde{\ell}_{L}$}
\end{picture}}
\end{center}
\end{minipage}
\hspace{-.7cm}
\begin{minipage}[h]{3cm}
\begin{center}
{\setlength{\unitlength}{1cm}
\begin{picture}(3,2.9)
\put(+1.5,.0){\includegraphics{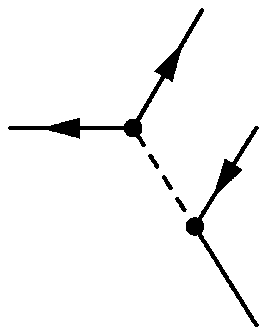}}
\put(1.1,1.3){$\stackrel{(-)}{\nu_{\ell}}$}
\put(.7,2.2){$\ell^{\pm}$}
\put(1.1,-.7){$\tilde{\chi}^0_{k}$}
\put(-1.4,1.4){$\tilde{\chi}^{\pm}_{i}$}
\put(-.2,.5){$\tilde{\nu}_{\ell}$}
\end{picture}}
\end{center}
\end{minipage}
\vspace{.2cm}
\begin{minipage}[h]{3cm}
\begin{center}
{\setlength{\unitlength}{1cm}
\begin{picture}(3,2.5)
\put(.8,-.1){\includegraphics{prowmq.ps}}
\put(2.6,-1){$\ell^-$}
\put(.2,1.2){$\tilde{\chi}^{0}_{i}$}
\put(2.6,.9){$\ell^{+}$}
\put(2.2,1.9){$\tilde{\chi}^0_{k}$}
\put(.8,.1){$Z^{0}$}
\end{picture}}
\end{center}
\end{minipage}
\hspace{2.4cm}
\vspace{.8cm}
\begin{minipage}[h]{3cm}
\begin{center}
{\setlength{\unitlength}{1cm}
\begin{picture}(3,2.9)
\put(+.3,+0){\includegraphics{proslmq.ps}}
\put(.5,2.1){$\ell^-$}
\put(.9,1.1){$\ell^{+}$}
\put(-1.5,1.4){$\tilde{\chi}^{0}_{i}$}
\put(.9,-.8){$\tilde{\chi}^0_{k}$}
\put(-.8,.5){$\tilde{\ell}_{L,R}$}
\end{picture}}
\end{center}
\end{minipage}
\hspace{.7cm}
\begin{minipage}[h]{3cm}
\begin{center}
{\setlength{\unitlength}{1cm}
\begin{picture}(3,2.9)
\put(+1.5,.0){\includegraphics{prosnmq.ps}}
\put(1.1,1.3){$\ell^-$}
\put(.7,2.2){$\ell^{+}$}
\put(1.1,-.7){$\tilde{\chi}^0_{k}$}
\put(-1.4,1.4){$\tilde{\chi}^{0}_{i}$}
\put(-.6,.5){$\tilde{\ell}_{L,R}$}
\end{picture}}
\end{center}
\end{minipage}
\begin{picture}(10,1)
\put(0,-.1){Figure 1b: 
Feynman diagrams for chargino decays
$\tilde{\chi}^{\pm}_{i}\to
\tilde{\chi}^0_{k} \ell^{\pm} \stackrel{(-)}{\nu_{\ell}}$ and} 
\put(.1,-.6){neutralino decays
$\tilde{\chi}^0_{i}\to
\tilde{\chi}^0_{k} \ell^+ \ell^-$.} 
\end{picture}
\end{figure}

\begin{figure}
\hspace{-.8cm}
\begin{minipage}{7cm}
\begin{picture}(7,7)
\put(0,0){\includegraphics{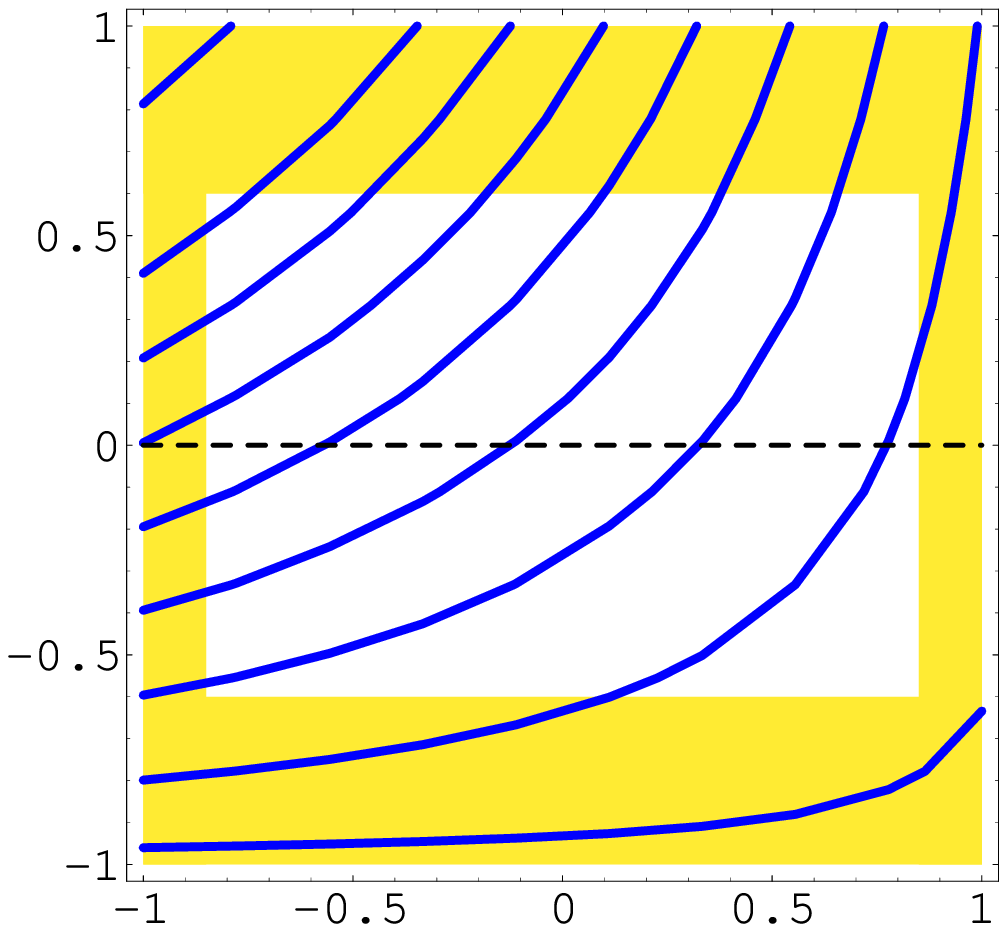}}
\put(1.3,5){a) Scenario~A1, $\tilde{\chi}^+_1 \tilde{\chi}^-_1$  }
\put(6.2,-1.2){ \small $ P_{e^-}$}
\put(-.3,4.6){\small $ P_{e^+}$}
\put(5.55,-.4){\tiny 10}
\put(4.65,.6){\tiny 50}
\put(3.8,1.2){\tiny 100}
\put(3.2,1.7){\tiny 150}
\put(2.55,2.1){\tiny 200}
\put(2.2,2.55){\tiny 300}
\put(1.9,2.95){\tiny 350}
\put(1.5,3.3){\tiny 400}
\put(1.3,3.7){\tiny 450}
\put(1.05,4.0){\tiny 500}
\end{picture}\par\vspace{.6cm}
\end{minipage}\hfill\hspace{.2cm}
\begin{minipage}{7cm}
\begin{picture}(7,7)
\put(0,0){\includegraphics{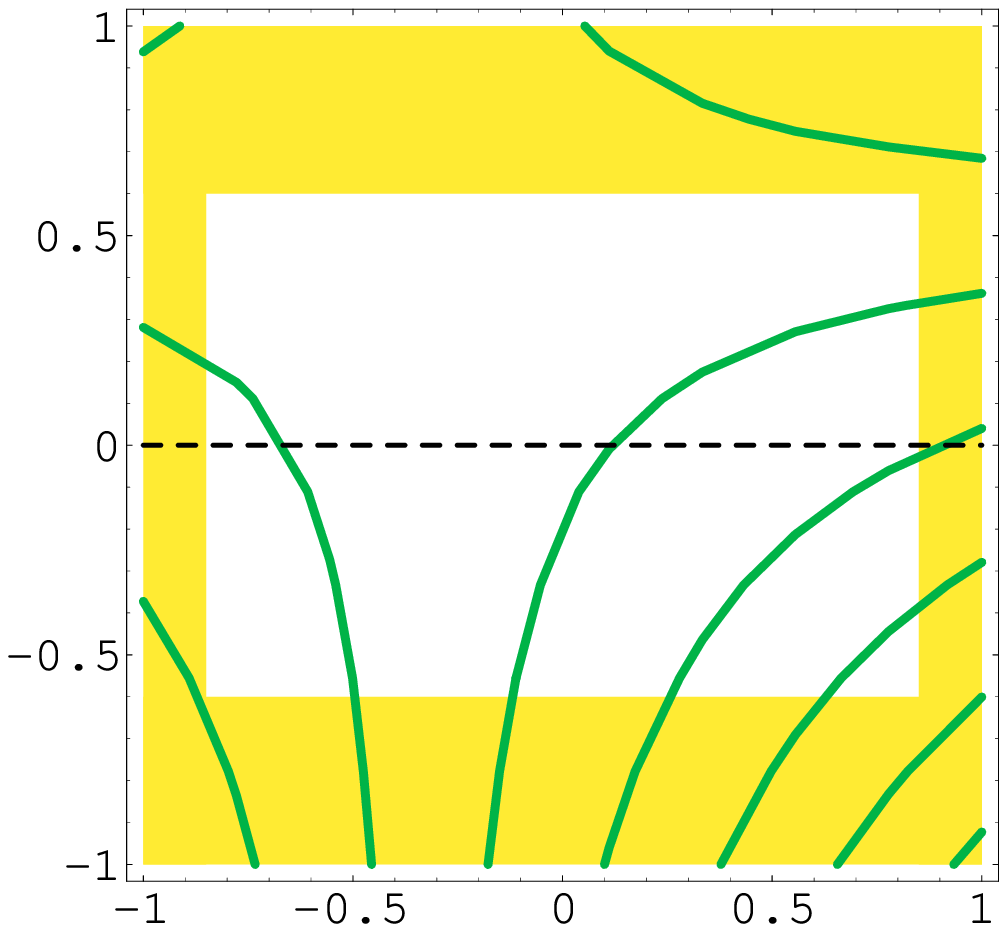}}
\put(1.3,5){b) Scenario~A1, $\tilde{\chi}^+_1 \tilde{\chi}^-_2$  }
\put(6.2,-1.2){\small $ P_{e^-}$}
\put(-.3,4.6){\small $ P_{e^+}$}
\put(1,4.2){\tiny 3}
\put(3.9,3.9){\tiny 2}
\put(1.6,2.3){\tiny 2}
\put(1.3,0.6){\tiny 1}
\put(3.6,2.2){\tiny 3}
\put(4.3,1.1){\tiny 4}
\put(4.8,0.55){\tiny 5}
\put(5.2,-.05){\tiny 6}
\put(5.7,-.4){\tiny 7}
\end{picture}\par\vspace{.2cm}
\hspace{-.5cm}
\end{minipage}
\hspace*{-.8cm}
\begin{minipage}{7cm}
\vspace{-1cm}
\begin{center}
\begin{picture}(7,7)
\put(0,0){\includegraphics{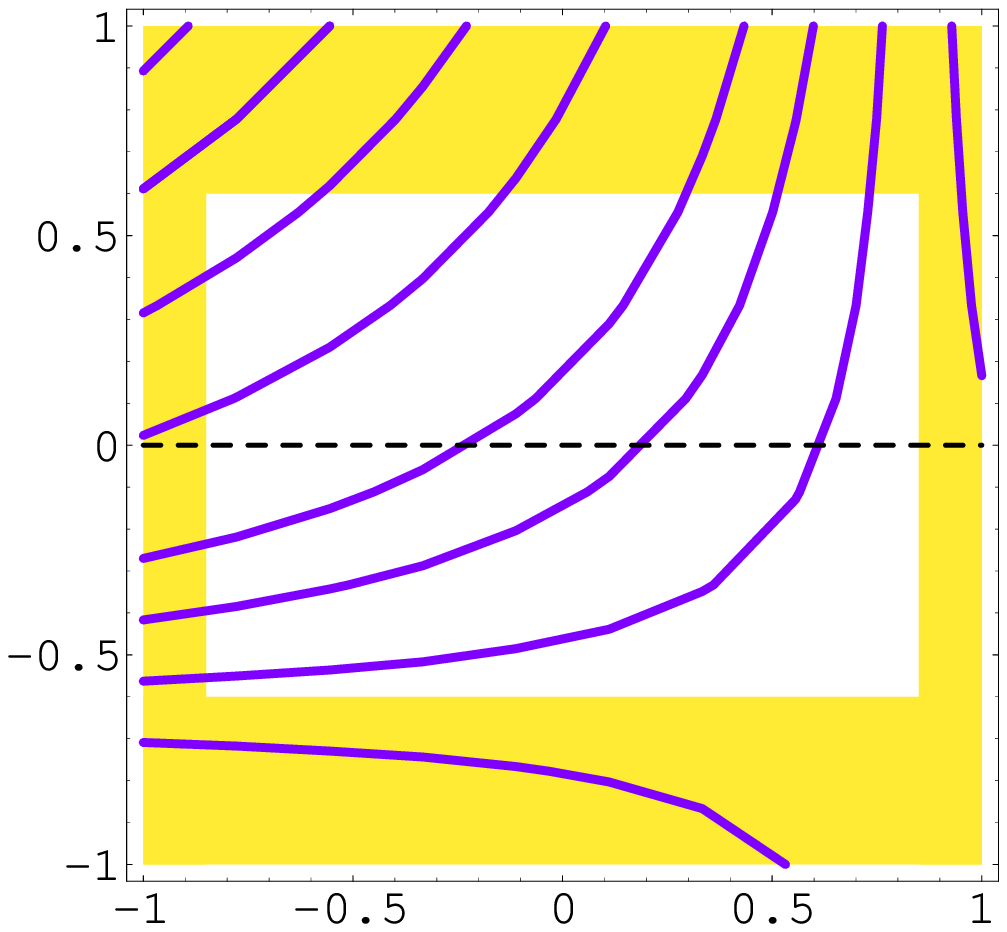}}
\put(1.3,5){c) Scenario~A1, $\tilde{\chi}^+_2 \tilde{\chi}^-_2$ }
\put(6.2,-1.2){\small $ P_{e^-}$}
\put(-.3,4.6){\small $ P_{e^+}$}
\put(2.4,-.3){\tiny 10}
\put(4.7,-.3){\tiny 20}
\put(5.5,2.3){\tiny 20}
\put(3.8,.6){\tiny 30}
\put(3.2,1.2){\tiny 40}
\put(2.8,1.65){\tiny 50}
\put(2.4,2.5){\tiny 70}
\put(1.8,3.1){\tiny 90}
\put(1.3,3.6){\tiny 110}
\put(1,4.1){\tiny 130}
\end{picture}\par\vspace{.6cm}
\end{center}
\end{minipage}\hfill\hspace{.2cm}
\begin{minipage}{7cm}
\begin{picture}(7,7)
\put(0,0){\includegraphics{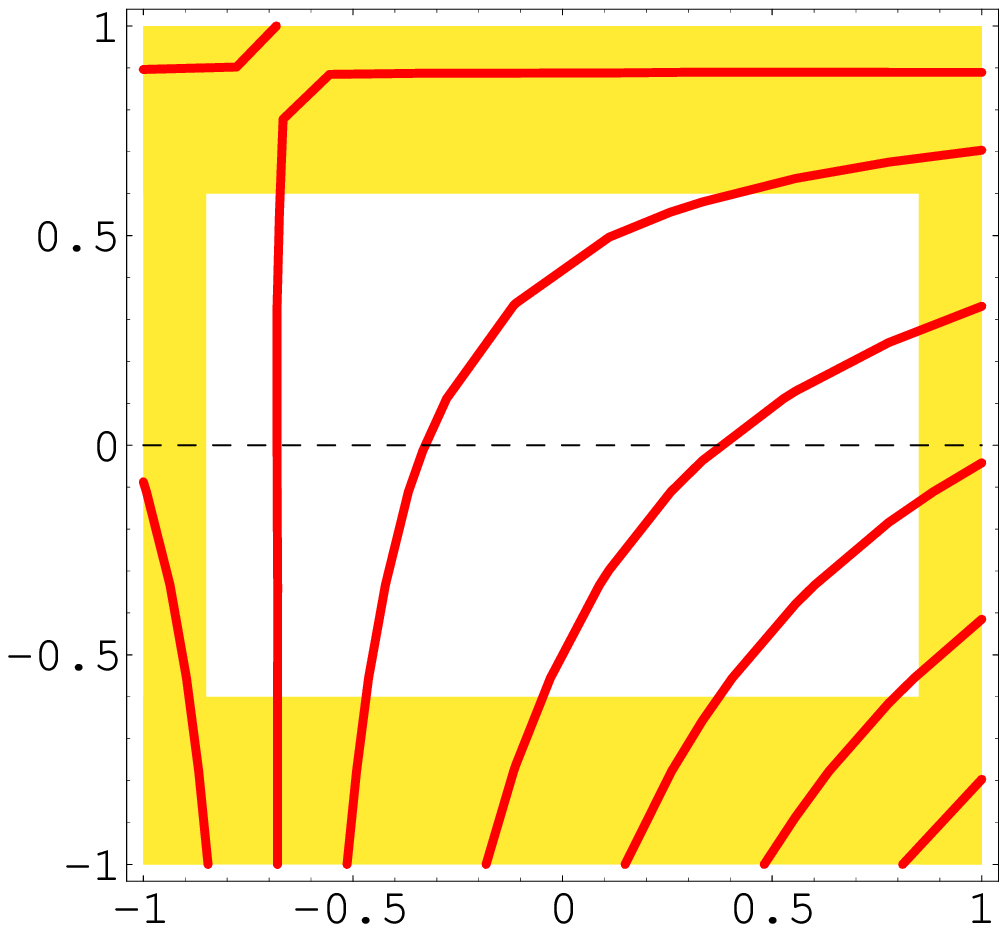}}
\put(1.3,5){d) Scenario~B, $\tilde{\chi}^+_1 \tilde{\chi}^-_2$ }
\put(6.2,-1.2){\small $ P_{e^-}$}
\put(-.3,4.6){\small $ P_{e^+}$}
\put(1,3.95){\tiny 2}
\put(1.4,3.45){\tiny 2}
\put(2.6,2.45){\tiny 3}
\put(1.3,0.6){\tiny 1}
\put(3.6,1.5){\tiny 5}
\put(4.3,.7){\tiny 7}
\put(5,.2){\tiny 9}
\put(5.4,-.3){\tiny 11}
\end{picture}\par\vspace{1cm}
\hspace{-.5cm}
\end{minipage}
\addtocounter{figure}{1}
\caption{Contour lines of cross section 
$\sigma(e^+e-\to \tilde{\chi}^+_i \tilde{\chi}^-_j)$ at 
$\sqrt{s}=m_{\tilde{\chi}^{+}_i}+m_{\tilde{\chi}^{-}_j}+10$~GeV. 
The longitudinal beam polarization for electrons (positrons) is denoted by
$P_{e^-}$ ($P_{e^+}$). The white region is for $|P_{e^-}|\le 85\%$,
$|P_{e^+}|\le 60\%$ (dashed line if only the electron beam is polarized).
\la{fig_2}}
\end{figure}

\begin{figure}
\hspace{-.8cm}
\begin{minipage}{7cm}
\begin{picture}(7,7)
\put(0,0){\includegraphics{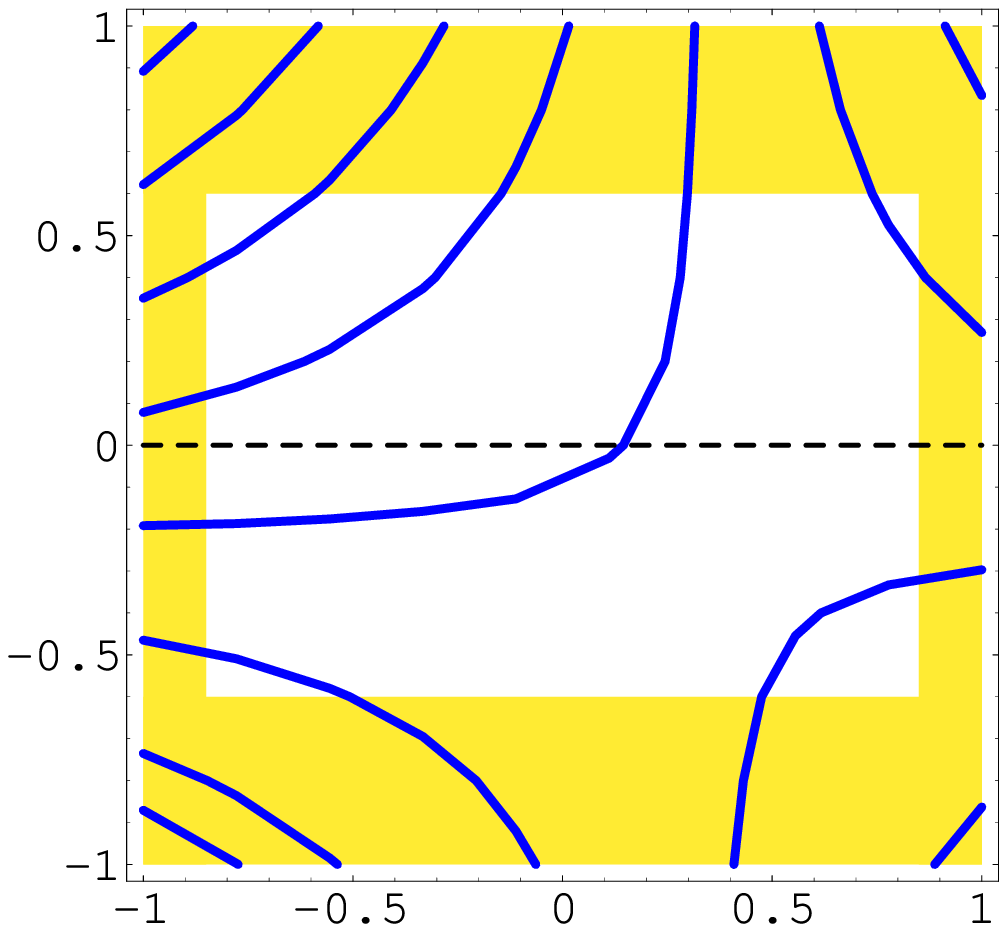}}
\put(1.3,5){a) Scenario~A1, $\tilde{\chi}^0_1 \tilde{\chi}^0_2$}
\put(6.2,-1.2){ \small $ P_{e^-}$}
\put(-.3,4.6){\small $ P_{e^+}$}
\put(1.1,-.3){\tiny 5}
\put(1.3,0){\tiny 10}
\put(1.8,.7){\tiny 20}
\put(4.6,1){\tiny 30}
\put(5.5,-.3){\tiny 40}
\put(4.1,2.1){\tiny 30}
\put(2.7,2.75){\tiny 40}
\put(2,3.3){\tiny 50}
\put(1.5,3.7){\tiny 60}
\put(1,4.05){\tiny 70}
\put(4.9,3.3){\tiny 20}
\put(5.5,4){\tiny 10}
\end{picture}\par\vspace{1cm}
\end{minipage}\hfill\hspace{.2cm}
\begin{minipage}{7cm}
\begin{picture}(7,7)
\put(0,0){\includegraphics{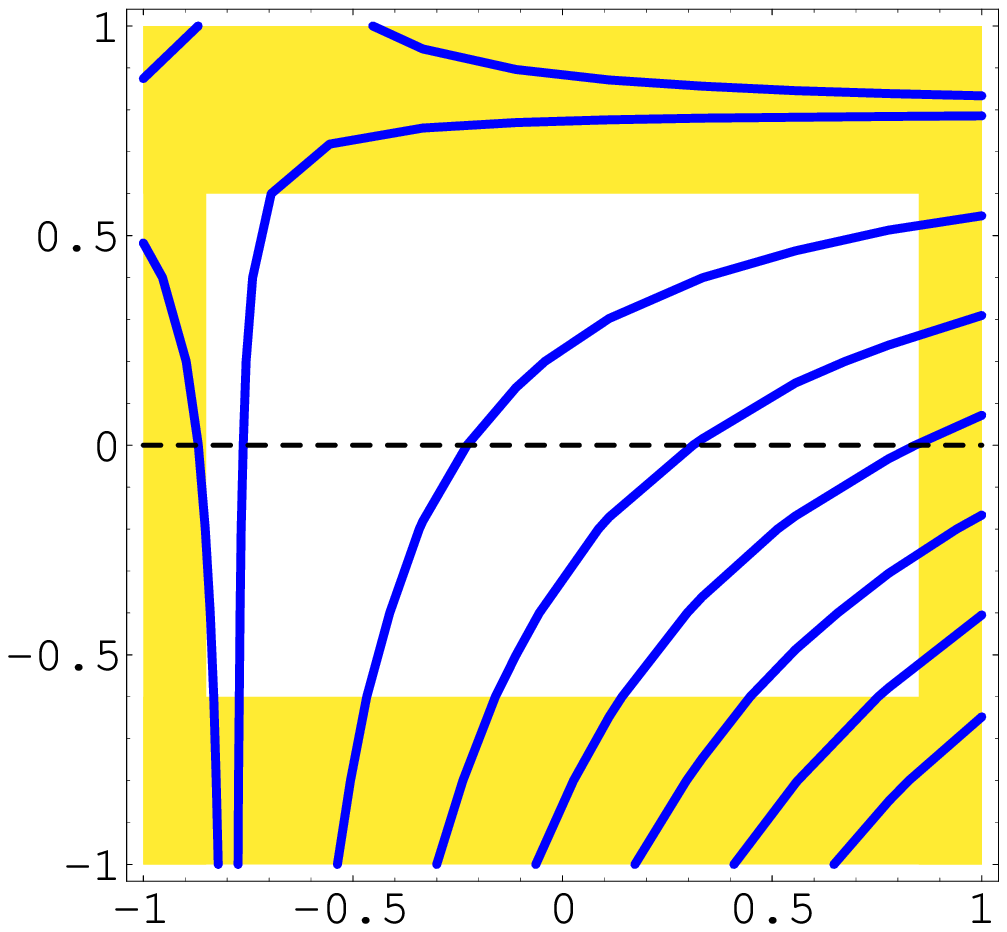}}
\put(1.3,5){b) Scenario~A2, $\tilde{\chi}^0_1 \tilde{\chi}^0_2$, 
}
\put(6.2,-1.2){\small $ P_{e^-}$}
\put(-.3,4.6){\small $ P_{e^+}$}
\put(1.1,4.1){\tiny 5}
\put(1,3.2){\tiny 4}
\put(2.1,4.2){\tiny 4}
\put(1.55,3.65){\tiny 5}
\put(5.25,0){\tiny 35}
\put(4.9,.3){\tiny 30}
\put(4.5,.7){\tiny 25}
\put(4,1.2){\tiny 20}
\put(3.55,1.7){\tiny 15}
\put(2.7,2.3){\tiny 10}
\end{picture}\par
\hspace{-.5cm}
\end{minipage}
\hspace*{-.8cm}
\begin{minipage}{7cm}
\vspace{-1cm}
\begin{center}
\begin{picture}(7,7)
\put(0,0){\includegraphics{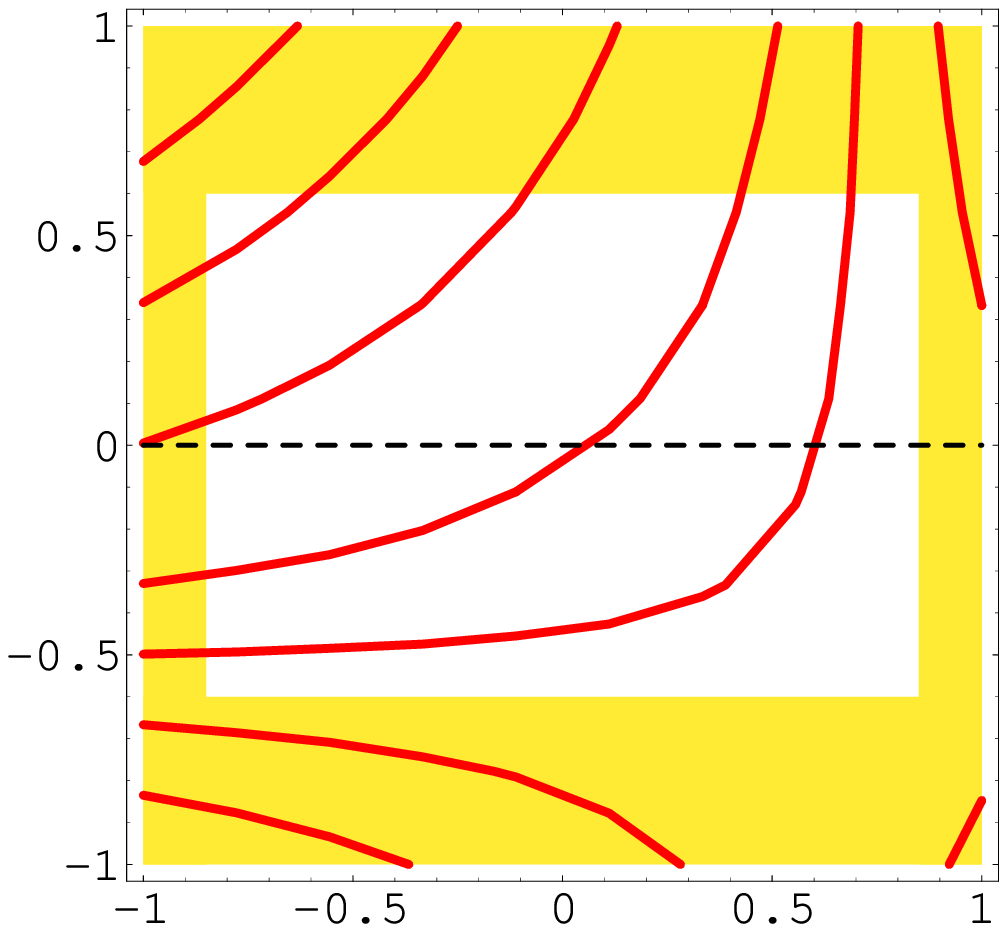}}
\put(1.3,5){c) Scenario~B, $\tilde{\chi}^0_1 \tilde{\chi}^0_2$}
\put(6.2,-1.2){\small $ P_{e^-}$}
\put(-.3,4.6){\small $ P_{e^+}$}
\put(5.5,3.4){\tiny 10}
\put(5.5,-.4){\tiny 15}
\put(3.5,0){\tiny 10}
\put(2,-.3){\tiny 5}
\put(4.6,1){\tiny 15}
\put(3.8,1.7){\tiny 20}
\put(2.8,2.7){\tiny 35}
\put(2,3.3){\tiny 45}
\put(1.5,3.9){\tiny 55}
\end{picture}\par\vspace{1.5cm}
\end{center}
\end{minipage}\hfill\hspace{.2cm}
\begin{minipage}{7cm}
\vspace{-1cm}
\begin{center}
\begin{picture}(7,7)
\put(0,0){\includegraphics{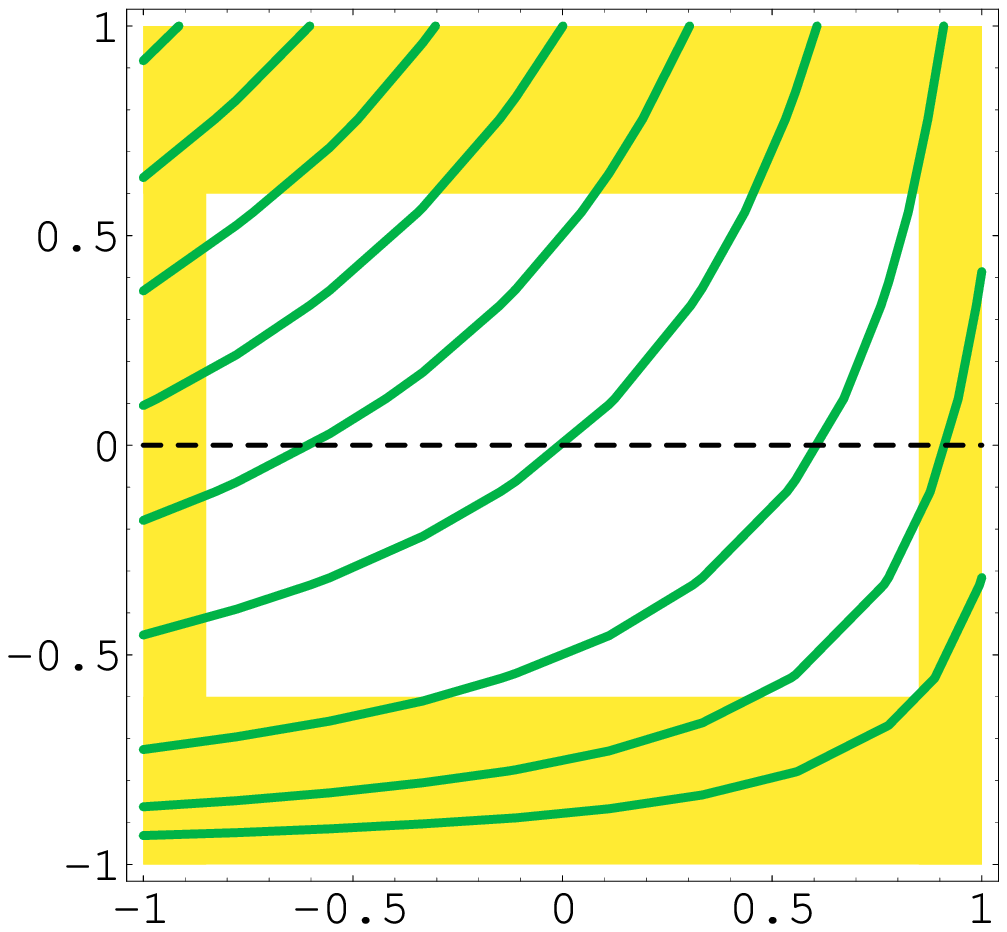}}
\put(1.3,5){d) Scenario~A1, $\tilde{\chi}^0_2 \tilde{\chi}^0_2$}
\put(6.2,-1.2){\small $ P_{e^-}$}
\put(-.3,4.6){\small $ P_{e^+}$}
\put(5.35,-.1){\tiny 5}
\put(4.9,.3){\tiny 10}
\put(4.3,.9){\tiny 20}
\put(3.5,1.7){\tiny 40}
\put(3,2.5){\tiny 60}
\put(2.4,3){\tiny 80}
\put(1.9,3.4){\tiny 100}
\put(1.5,3.85){\tiny 120}
\put(1,4.2){\tiny 140}
\end{picture}\par\vspace{1.5cm}
\end{center}
\end{minipage}
\caption{Contour lines of cross section 
$\sigma(e^+e-\to \tilde{\chi}^0_i \tilde{\chi}^0_j)$ at 
$\sqrt{s}=m_{\tilde{\chi}^{0}_i}+m_{\tilde{\chi}^{0}_j}+30$~GeV 
in scenario~A1. 
The longitudinal beam polarization for electrons (positrons) is denoted by
$P_{e^-}$ ($P_{e^+}$). The white region is for $|P_{e^-}|\le 85\%$,
$|P_{e^+}|\le 60\%$ (dashed line if only the electron beam is polarized).
\la{fig_3}}
\end{figure}

\begin{figure}
\hspace{-.8cm}
\begin{minipage}{7cm}
\begin{picture}(7,7)
\put(0,0){\includegraphics{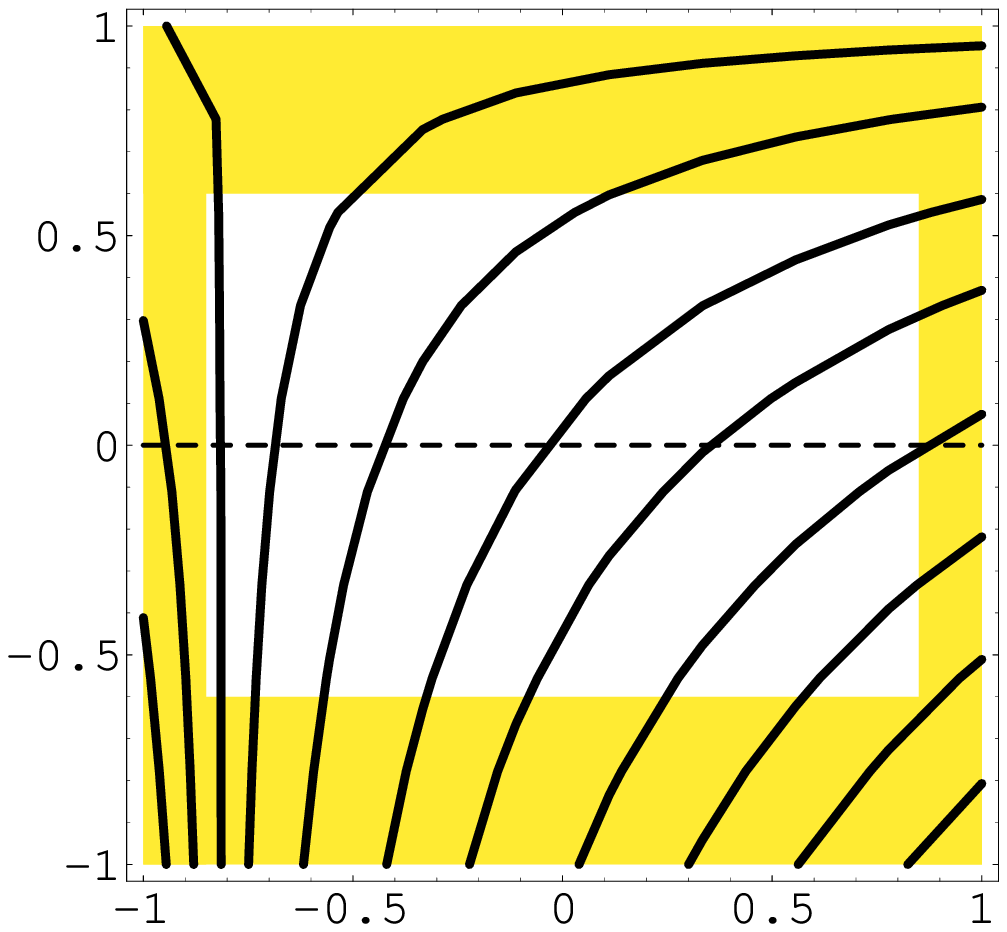}}
\put(.9,5){a) Scenario~A1, $e^+ e^-\to \tilde{\chi}^0_1 \tilde{\chi}^0_3$}
\put(6.2,-1.2){ \small $ P_{e^-}$}
\put(-.3,4.6){\small $ P_{e^+}$}
\put(.9,1.1){\tiny 1}
\put(1.1,2.5){\tiny 2}
\put(1.4,4.1){\tiny 3}
\put(1.75,3.4){\tiny 4}
\put(2.5,2.8){\tiny 6}
\put(3.2,2.2){\tiny 9}
\put(3.65,1.65){\tiny 12}
\put(4.2,1.05){\tiny 16}
\put(4.7,.6){\tiny 20}
\put(5.1,.25){\tiny 24}
\put(5.5,-.25){\tiny 28}
\end{picture}\par\vspace{1cm}
\end{minipage}\hfill\hspace{.2cm}
\begin{minipage}{7cm}
\begin{picture}(7,7)
\put(0,0){\includegraphics{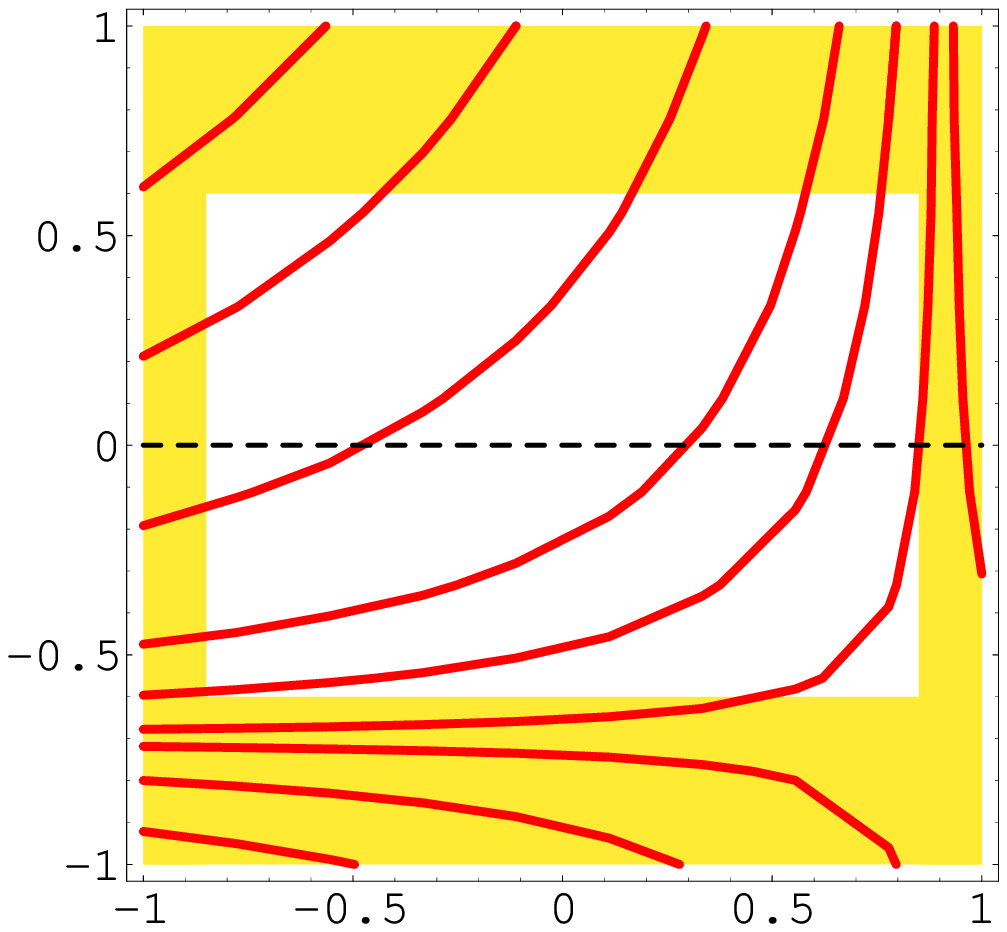}}
\put(.9,5){b) Scenario~A1, $e^+ e^-\to \tilde{\chi}^0_2 \tilde{\chi}^0_3$}
\put(6.2,-1.2){\small $ P_{e^-}$}
\put(-.3,4.6){\small $ P_{e^+}$}
\put(1.3,3.6){\tiny 40}
\put(2.1,3){\tiny 30}
\put(3,2.1){\tiny 20}
\put(3.6,1.2){\tiny 13}
\put(4.3,.8){\tiny 10}
\put(5.15,.35){\tiny 8}
\put(5.2,-.2){\tiny 7}
\put(4,-.4){\tiny 5}
\put(2.2,-.5){\tiny 2}
\put(5.8,.9){\tiny 7}
\end{picture}\par
\hspace{-.5cm}
\end{minipage}
\hspace*{-.8cm}
\caption{Contour lines of cross section 
$\sigma(e^+ e^-\to\tilde{\chi}^0_i\tilde{\chi}^0_j)$ in fb 
at $\sqrt{s}=(m_{\tilde{\chi}^0_1}+m_{\tilde{\chi}^0_2})+30$~GeV
in scenario~A1. 
The longitudinal beam polarization for electrons
(positrons) is denoted by $P_{e^-}$ ($P_{e^+}$).
The white region is for $|P_{e^-}|<85\%$, $|P_{e^+}|<60\%$
(dashed--line if only  electron beam polarized). }
\hspace{-1cm}
\begin{minipage}{7cm}
\vspace{1.2cm}
\begin{picture}(7,5)
\put(-.1,0){\includegraphics{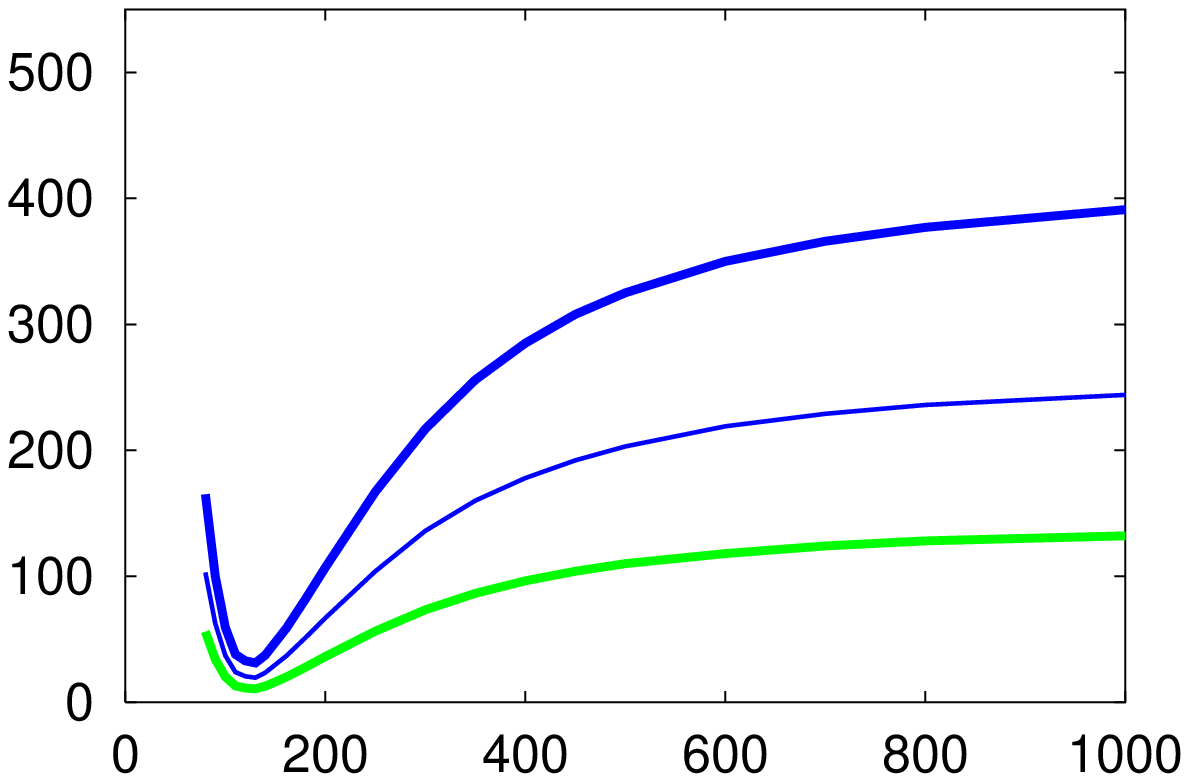}}
\put(1.8,5.1){a) $\sqrt{s}=2 m_{\tilde{\chi}^{\pm}_1}+10$~GeV}
\put(5.8,-.2){\small $ m_{\tilde{\nu}_e}${\small /GeV}}
\put(-.1,5.4){\small $ \sigma_e${\small /fb}}
\put(5.9,1.8){\small $(00)$}
\put(5.8,2.7){\small $(-0)$}
\put(5.8,3.9){\small $(-+)$}
\end{picture}\par\vspace{.2cm}
\end{minipage}\hfill\hspace{.2cm}
\begin{minipage}{7cm}
\vspace{-1cm}
{\parbox{6.5cm}{\small Figure 5:
Cross section $\sigma_e=\sigma(e^+e^-\to\tilde{\chi}^+_1\tilde{\chi}^-_1)
\times BR(\tilde{\chi}^-_1\to \tilde{\chi}^0_1 e^- \bar{\nu}_e$
in fb at $\sqrt{s}=2m_{\tilde{\chi}^{\pm}_1}+10$~GeV 
as function of 
$m_{\tilde{\nu}_e}$ for unpolarized beams (00), 
only the electron beam polarized 
$P_-^3=-85\%$ $(-0)$ and both beams polarized $P_-^3=-85\%$, 
$P_+^3=+60\%$ $(-+)$. Other parameters as in scenario~A1. }}
\end{minipage}
\end{figure}
\begin{figure}[t]
\hspace{-.8cm}
\begin{minipage}{7cm}
\vspace{-1.5cm}
\begin{picture}(7,7)
\put(0,0){\includegraphics{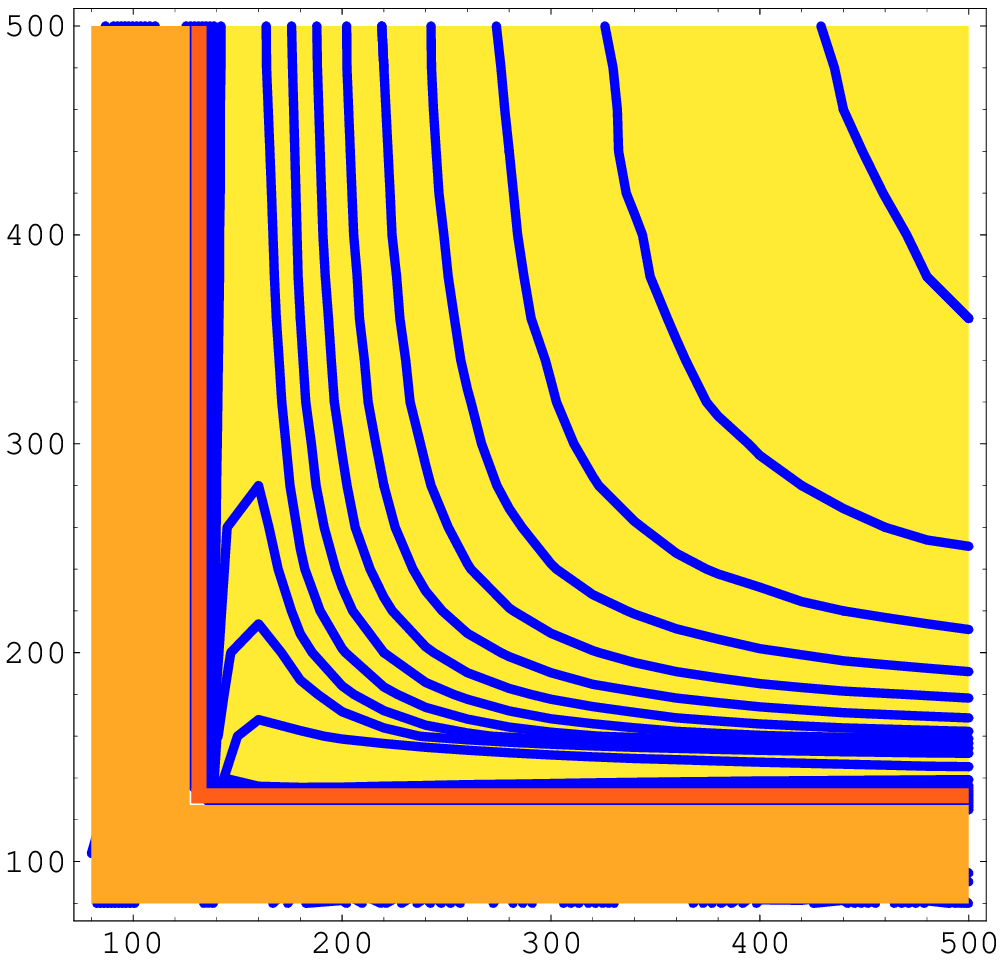}}
\put(1.2,5){a) $\sqrt{s}=2 m_{\tilde{\chi}^{\pm}_1}+10$~GeV}
\put(1.7,-1.3){\tiny $A_{FB}=10,12,14,\ldots,30,34$[\%]}
\put(6,-1.3){\small $ m_{\tilde{\nu}}$}
\put(-.4,4.3){\small $ m_{\tilde{e}_L}$}
\put(5.3,4){\tiny 10}
\put(4.1,3){\tiny 12}
\put(3.45,2.4){\tiny 14}
\put(3.,1.95){\tiny 16}
\put(1.35,1.9){\tiny 28}
\put(1.4,1.05){\tiny 30}
\put(1.4,.45){\tiny 34}
\end{picture}\par\vspace{1cm}
\end{minipage}\hfill\hspace{.2cm}
\begin{minipage}{7cm}
\vspace{-1.5cm}
\begin{picture}(7,7)
\put(0,-.4){\includegraphics{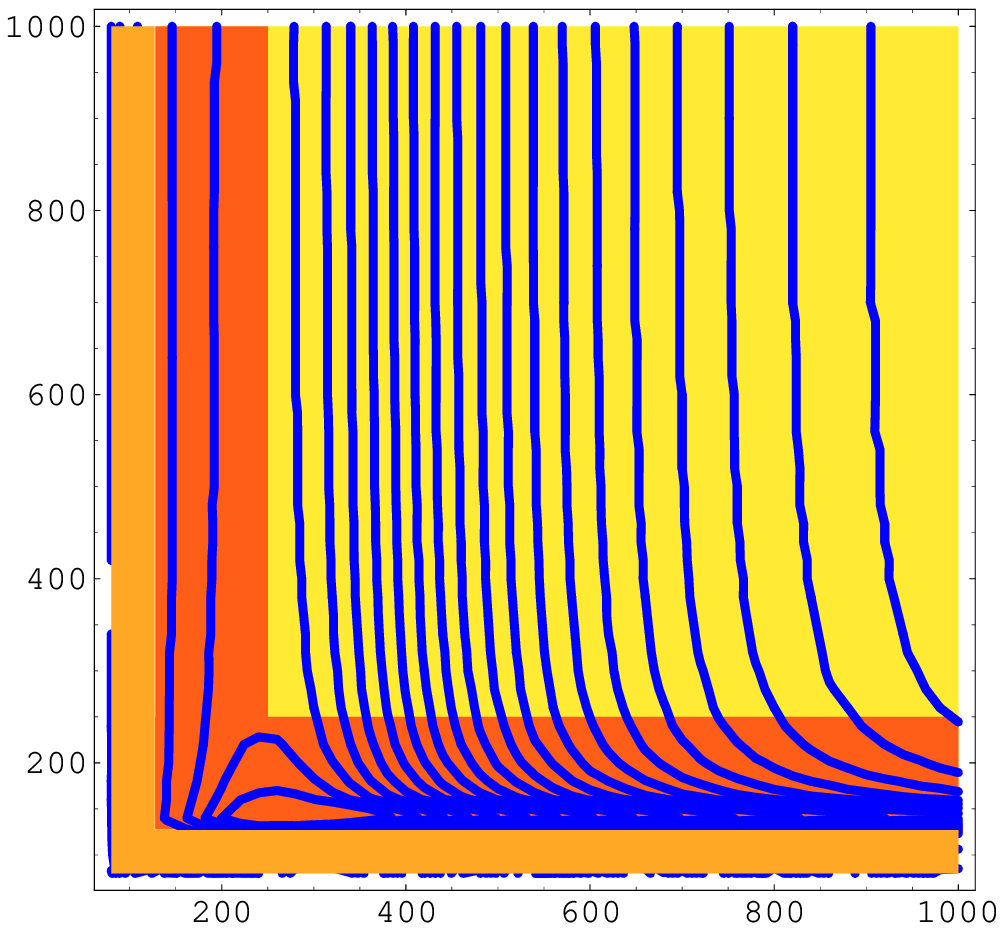}}
\put(1.9,4.8){b) $\sqrt{s}=500$~GeV}
\put(1.7,-1.6){\tiny $A_{FB}=12,14,16,\ldots,48,50$[\%]}
\put(6.3,-1.6){\small $ m_{\tilde{\nu}}$}
\put(-.4,4.1){\small $ m_{\tilde{e}_L}$}
\put(5.6,3.5){\tiny 12}
\put(5.1,3.3){\tiny 14}
\put(4.7,3.1){\tiny $\ldots$}
\put(1.5,-.05){\tiny 48}
\put(1.55,-.4){\tiny 50}
\put(1.4,3.5){\tiny 46}
\put(1.1,4){\tiny 44}
\end{picture}\par\vspace{1cm}
\hspace{-.5cm}
\end{minipage}
\addtocounter{figure}{1}
\caption{Contour lines of the forward--backward asymmetry of the decay 
electron $A_{FB}/\%$ of
$e^+ e^-\to \tilde{\chi}^+_1 \tilde{\chi}^-_1$, 
$\tilde{\chi}^-_1\to \tilde{\chi}^0_1 e^- \bar{\nu}$
at a) $\sqrt{s}=2 m_{\tilde{\chi}^{\pm}_1}+10$~GeV and b) 
$\sqrt{s}=500$~GeV, as a function of
$m_{\tilde{e}_L}$ and $m_{\tilde{\nu}}$ for
$P_{e^-}=-85\%$, $P_{e^+}=+60\%$, the other parameters as in 
scenario~A1.
The light covered region is dominated by the two--body decay 
$\tilde{\chi}^-_1 \to \tilde{e}_L \bar{\nu}_e$ or
$\tilde{\chi}^-_1 \to \tilde{\nu}_e e^-$. Outside 
the red coloured region
direct production of $\tilde{e}_L$ or $\tilde{\nu}_e$ is not possible,
$\sqrt{s}/2>m_{\tilde{e}_L},m_{\tilde{\nu}_e}$. \la{fig_5}}
\end{figure}

\begin{figure}[t]
\hspace{-.8cm}
\begin{minipage}{7cm}
\begin{picture}(7,7)
\put(0,0){\includegraphics{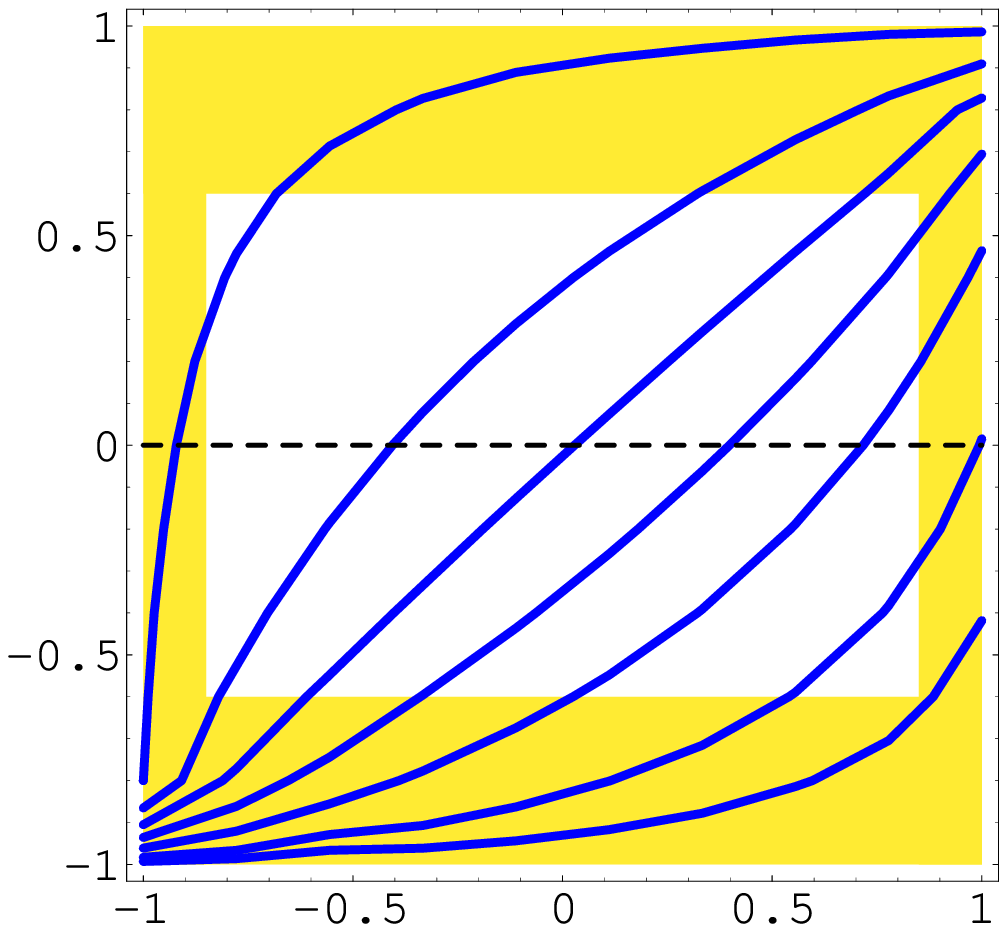}}
\put(.6,4.9){a) $m_{\tilde{e}_L}=176$~GeV, $m_{\tilde{e}_R}=132$~GeV}
\put(1.2,-1.2){\tiny $A_{FB}=-12,-8,-4,0,4,8,10$[\%]}
\put(6,-1.2){\small $ P_{e^-}$}
\put(0,4.1){\small $ P_{e^+}$}
\put(3.8,1.6){\tiny 0}
\put(4.1,.3){\tiny 8\%}
\put(3.6,.8){\tiny 4\%}
\put(4.5,0){\tiny 10\%}
\put(3.,2.1){\tiny $-4\%$}
\put(2.5,2.7){\tiny $-8\%$}
\put(1.2,3.8){\tiny $-12\%$}
\end{picture}\par\vspace{1cm}
\end{minipage}\hfill\hspace{.2cm}
\begin{minipage}{7cm}
\vspace*{-.5cm}
\begin{picture}(7,7)
\put(0,0){\includegraphics{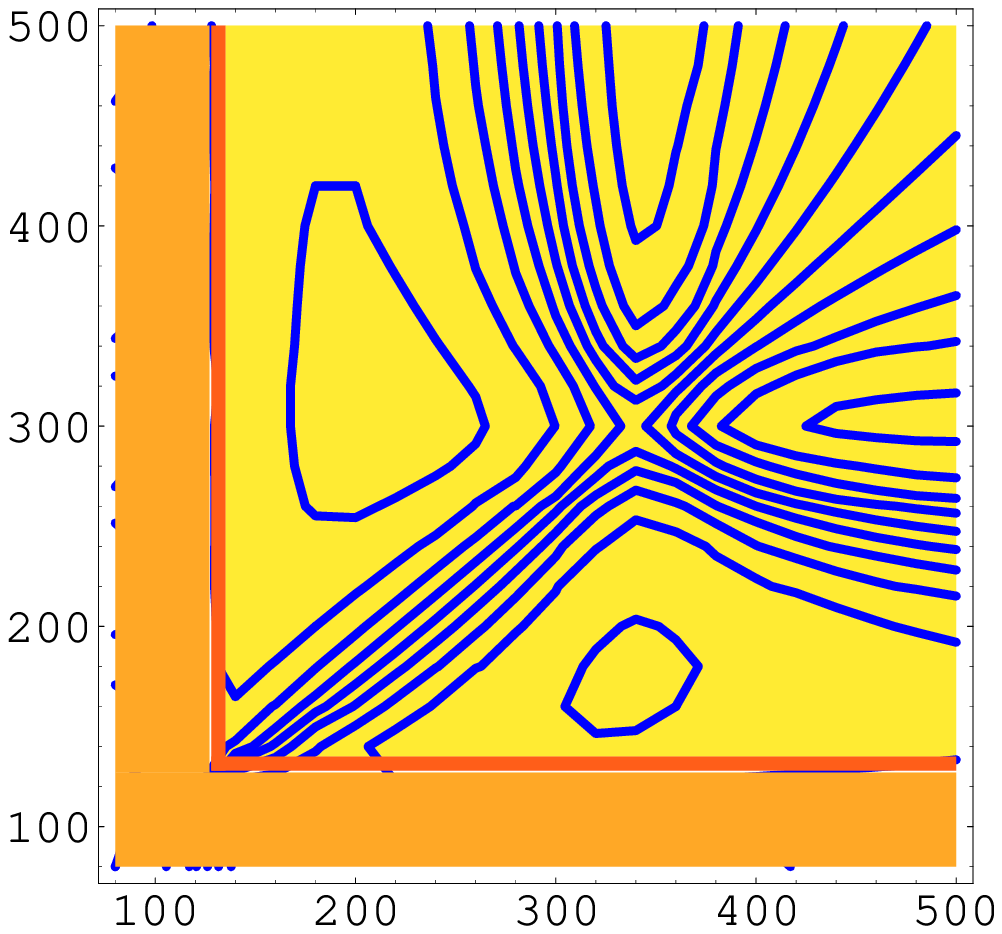}}
\put(.6,4.9){b) $P_{e^-}=-85\%$, $P_{e^+}=+60\%$}
\put(5.5,-1.2){\small $ m_{\tilde{e}_L}$}
\put(-.3,4.2){\small $ m_{\tilde{e}_R}$}
\put(1.2,-1.2){\tiny $A_{FB}=-20,-15,\ldots,15,19,23$[\%]}
\put(2.2,2.3){\tiny $19$}
\put(2.1,4.2){\tiny $15$}
\put(3.7,4.35){\tiny $-20$}
\put(5.1,4.35){\tiny $-5$}
\put(5.4,3.9){\tiny 0}
\put(5.4,3.25){\tiny $5$}
\put(5.25,2.35){\tiny $19$}
\put(5.5,2.1){\tiny $23$}
\put(3.6,1){\tiny $-15$}
\put(3.5,.5){\tiny $-20$}
\end{picture}\par
\hspace{-.5cm}
\end{minipage}
\hspace*{-.8cm}
\begin{minipage}{7cm}
\vspace{-1cm}
\begin{center}
\begin{picture}(7,7)
\put(0,0){\includegraphics{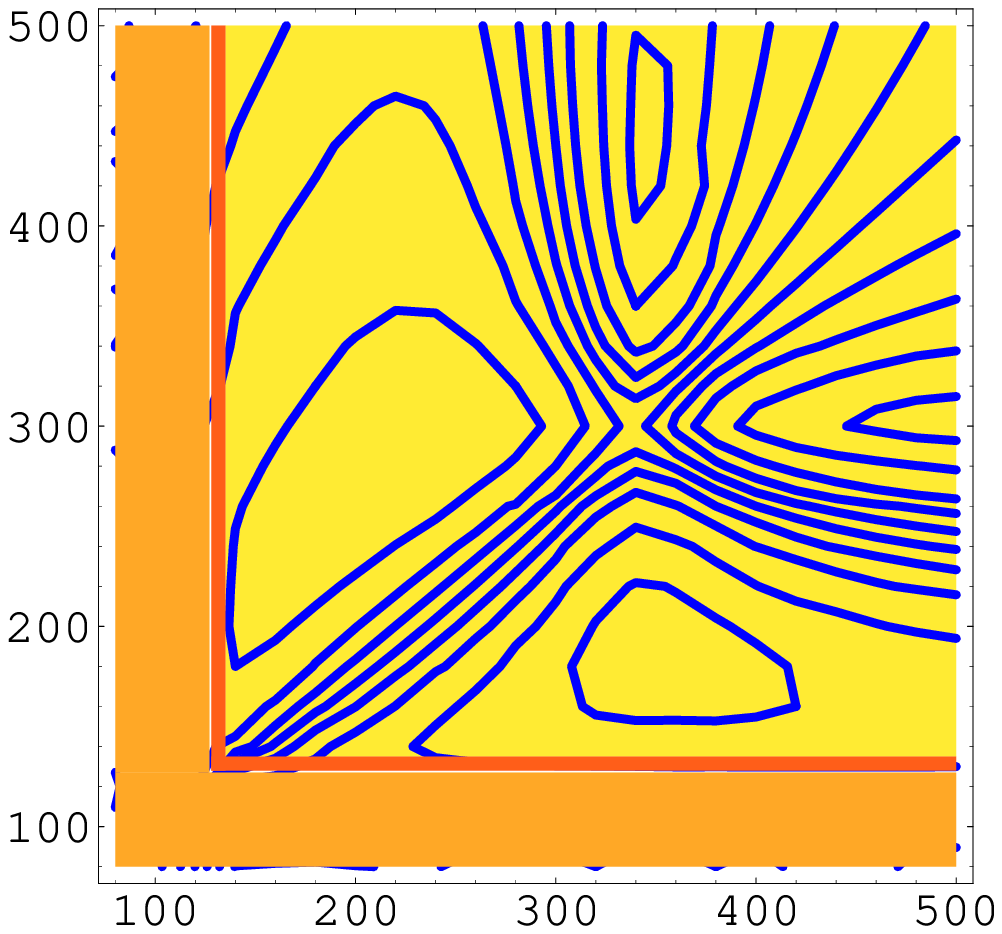}}
\put(.6,4.9){c) $P_{e^-}=+85\%$, $P_{e^+}=-60\%$}
\put(5.5,-1.2){\small $ m_{\tilde{e}_L}$}
\put(-.3,4.2){\small $ m_{\tilde{e}_R}$}
\put(1.1,-1.2){\tiny $A_{FB}=-23,-20,-15,\ldots,15,18$[\%]}
\put(3.7,.75){\tiny $18$}
\put(5.3,.5){\tiny 15}
\put(4.6,2.05){\tiny $-20$}
\put(5.45,2.06){\tiny $-23$}
\put(5.6,4.1){\tiny 0}
\put(5.4,3.3){\tiny $-5$}
\put(5.1,4.25){\tiny 5}
\put(3.85,4){\tiny 18}
\put(4,4.4){\tiny 15}
\put(2.4,4.25){\tiny $-5$}
\put(1.65,4.25){\tiny $-5$}
\put(2,3.5){\tiny $-10$}
\put(2.1,2.4){\tiny $-15$}
\end{picture}\par\vspace{.6cm}
\end{center}
\end{minipage}\hfill\hspace{.2cm}
\begin{minipage}{7cm}
\vspace{1.4cm}
{\parbox{6.5cm}{\small Figure 7:
Contour lines of the forward--backward
asymmetry of the decay electron 
$A_{FB}$/\% of
$e^+ e^-\to\tilde{\chi}^0_1\tilde{\chi}^0_2, \tilde{\chi}^0_2\to 
\tilde{\chi}^0_1 e^+ e^-$
at $\sqrt{s}=(m_{\tilde{\chi}^0_1}+m_{\tilde{\chi}^0_2})+30$~GeV
in scenario~A1 as a function of a) $P_{e^-}$ and $P_{e^+}$, 
b) $m_{\tilde{e}_L}$ and $m_{\tilde{e}_R}$, for
$P_{e^-}=-85\%$, $P_{e^+}=+60\%$, and c)  $m_{\tilde{e}_L}$ and 
$m_{\tilde{e}_R}$ for $P_{e^-}=+85\%$, $P_{e^+}=+60\%$.
The light covered region is dominated by the two--body decay 
$\tilde{\chi}^0_2 \to \tilde{e}_{L,R} e$. Outside the red coloured region
direct production of $\tilde{e}_L$ or
$\tilde{e}_R$ is not possible, 
$\sqrt{s}/2>m_{\tilde{e}_{L,R}}$.} }
\end{minipage}
\end{figure}

\begin{figure}
\hspace{-.9cm}
\begin{minipage}{7cm}
\begin{picture}(7,5)
\put(0,0){\includegraphics{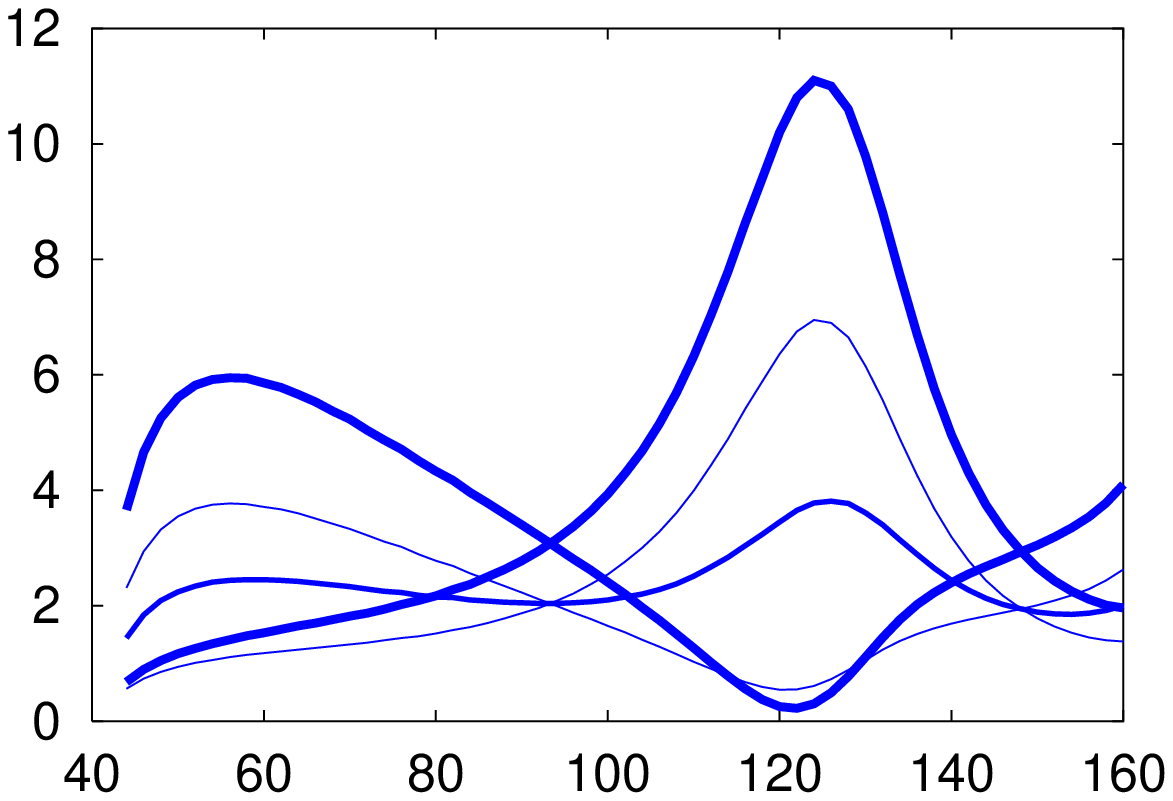}}
\put(1,5.2){\small a) $m_{\tilde{e}_L}=176$~GeV, $m_{\tilde{e}_R}=161$~GeV}
\put(5.8,-.3){ \small $M_1${\small /GeV}}
\put(-.4,5.2){\small $ \sigma_e${\small /fb}}
\put(4.7,2.1){\small $(00)$}
\put(4.6,3.3){\small $(+0)$}
\put(5.5,4){\small $(+-)$}
\put(1.1,2.1){\small $(-0)$}
\put(1.1,2.9){\small $(-+)$}
\end{picture}\par
\end{minipage}\hfill\hspace{.2cm}
\begin{minipage}{7cm}
\begin{picture}(7,5)
\put(0,0){\includegraphics{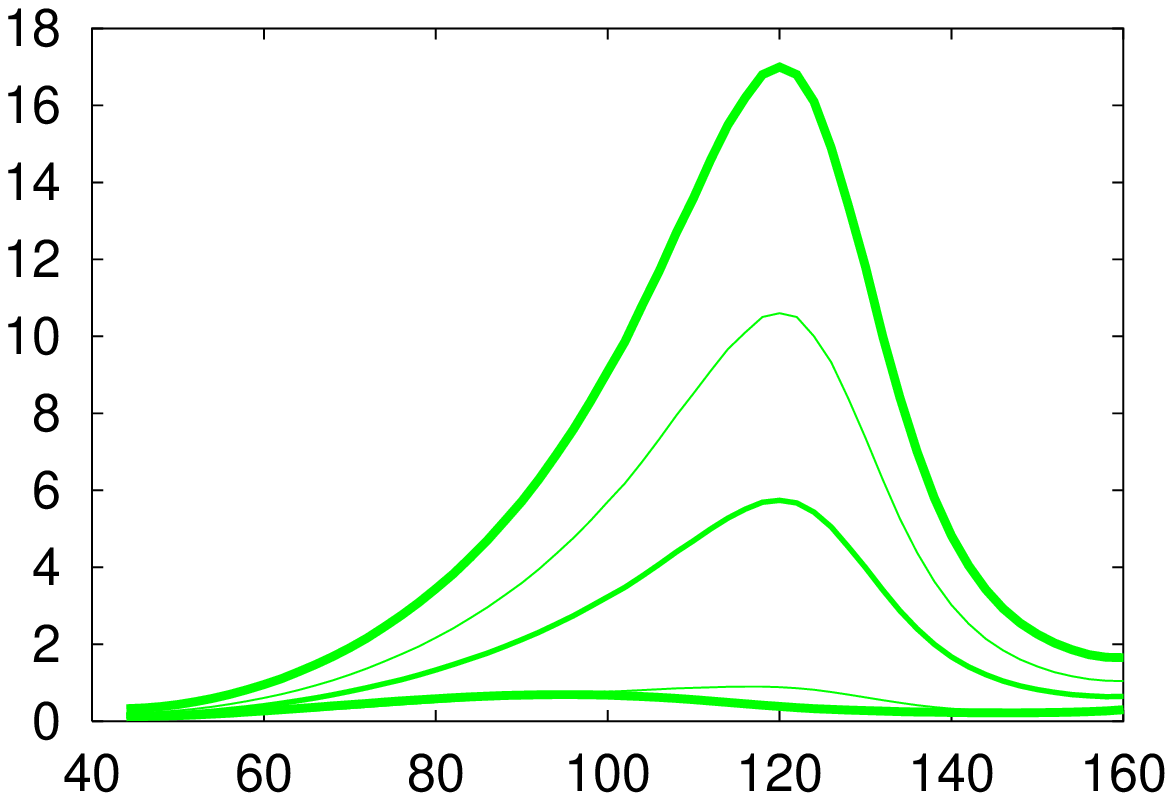}}
\put(1,5.2){\small b) $m_{\tilde{e}_L}=500$~GeV, $m_{\tilde{e}_R}=161$~GeV}
\put(5.8,-.3){\small $ M_1${\small /GeV}}
\put(-.2,5.2){\small $ \sigma_e$ {\small /fb}}
\put(3.5,1){\small $(-+)\approx (-0)$}
\put(4.4,2.1){\small $(00)$}
\put(4.4,3.3){\small $(+0)$}
\put(5.3,4.2){\small $(+-)$}
\end{picture}\par
\end{minipage}
\addtocounter{figure}{1}
\caption{Cross sections $\sigma_e$ at 
$\sqrt{s}=m_{\tilde{\chi}^0_1}+m_{\tilde{\chi}^0_2}+30$~GeV as function of 
gaugino parameter $M_1$ for unpolarized beams (00), 
for only electron beam polarized 
$(-0)$, $(+0)$ with $P_{e^-}=\pm 85\%$ and for both beams 
polarized $(-+)$, $(+-)$ 
with $P_-=\mp 85\%$, $P_{e^+}=\pm 60\%$. The slepton masses are  a) 
$m_{\tilde{e}_L}=176$~GeV, $m_{\tilde{e}_R}=161$~GeV, and b) 
$m_{\tilde{e}_L}=500$~GeV, $m_{\tilde{e}_R}=161$~GeV; the
other SUSY parameters as in scenario~A1. \la{fig_8}}
\end{figure}

\begin{figure}[t]
\hspace{-.9cm}
\begin{minipage}{7cm}
\begin{picture}(7,5)
\put(-.2,0){\includegraphics{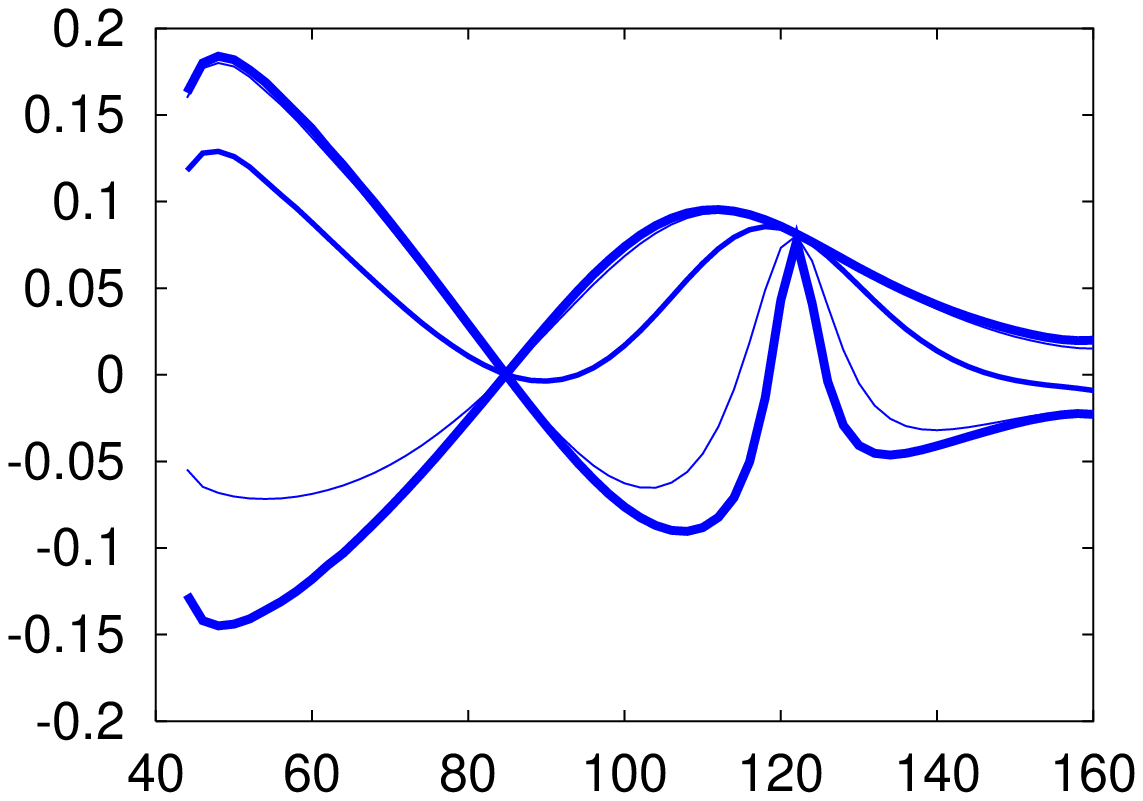}}
\put(1,5.2){\small a) $m_{\tilde{e}_L}=176$~GeV, $m_{\tilde{e}_R}=161$~GeV}
\put(5.5,-.3){ \small $ M_1${\small /GeV}}
\put(-.2,5.2){\small $ A_{FB}$}
\put(1.9,4.4){\small $(-+)\approx (-0)$}
\put(4,1.4){\small $(-+)$}
\put(3.6,2.4){\small $(-0)$}
\put(4.5,4){\small $(+-)\approx (+0)$}
\put(1.1,3.5){\small $(00)$}
\put(1.2,2.2){\small $(+0)$}
\put(1.9,1){\small $(+-)$}
\end{picture}\par\vspace{1cm}
\end{minipage}\hfill\hspace{.2cm}
\begin{minipage}{7cm}
\begin{picture}(7,5)
\put(-.2,0){\includegraphics{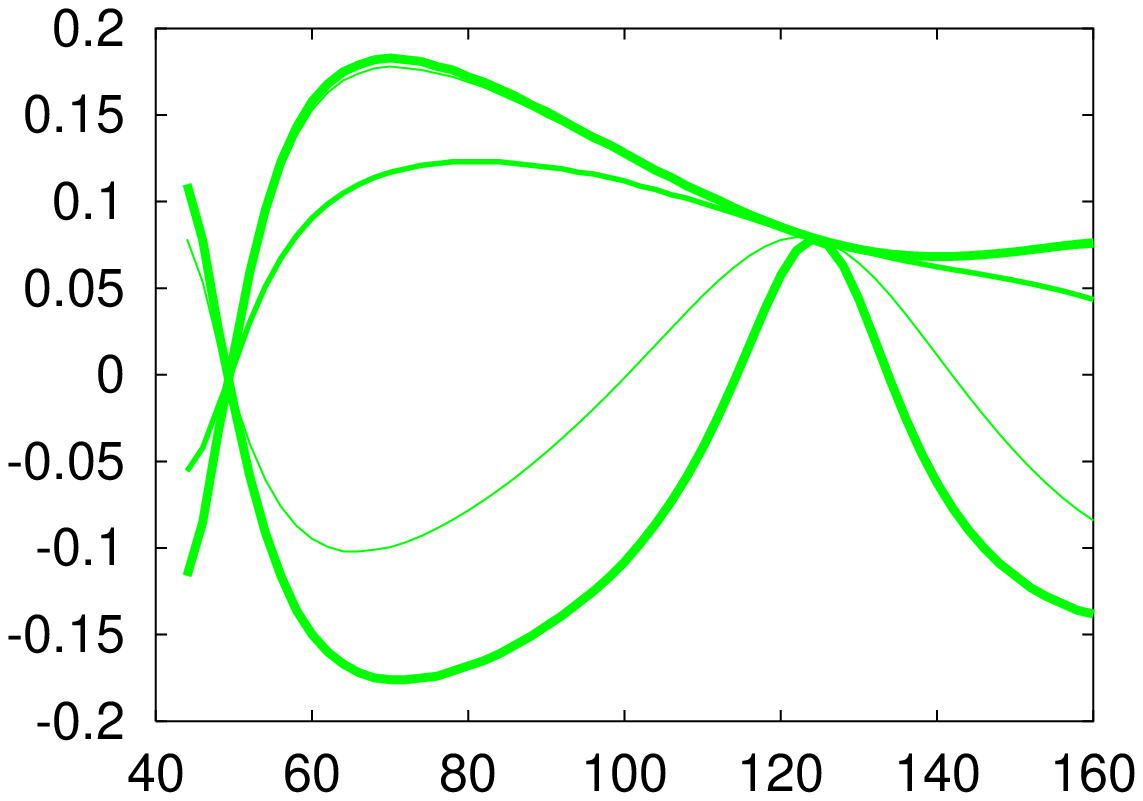}}
\put(1,5.2){\small b) $m_{\tilde{e}_L}=500$~GeV, $m_{\tilde{e}_R}=161$~GeV}
\put(5.5,-.3){\small $ M_1/${\small ~GeV}}
\put(0,5.2){\small $ A_{FB}$}
\put(3.5,4.4){\small $(+-)\approx (+0)$}
\put(2.1,3.4){\small $(00)$}
\put(2,2){\small $(-0)$}
\put(2.1,1.1){\small $(-+)$}
\end{picture}\par\hspace{-.5cm}\vspace{+.6cm}
\end{minipage}
\vspace{-1cm}
\caption{Forward--backward asymmetries $A_{FB}$ 
of the decay electron of 
$e^+ e^-\to \tilde{\chi}^0_1 \tilde{\chi}^0_2$, 
$\tilde{\chi}^0_2\to \tilde{\chi}^0_1 e^+ e^-$ at 
$\sqrt{s}=m_{\tilde{\chi}^0_1}+m_{\tilde{\chi}^0_2}+30$~GeV, as
a function of the gaugino parameter $M_1$
for a) $m_{\tilde{e}_L}=176$~GeV, $m_{\tilde{e}_R}=161$~GeV 
and b) $m_{\tilde{e}_L}=500$~GeV, $m_{\tilde{e}_R}=161$~GeV
for unpolarized beams (00), only 
electron beam polarized $(-0)$, $(+0)$ and both beams polarized
$(-+)$, $(+-)$; the other SUSY parameters as in scenario~A1. 
\la{fig_9}}
\end{figure}

\begin{figure}[t]
\hspace*{-.8cm}
\begin{minipage}{7cm}
\begin{picture}(7,5)
\put(-.2,0){\includegraphics{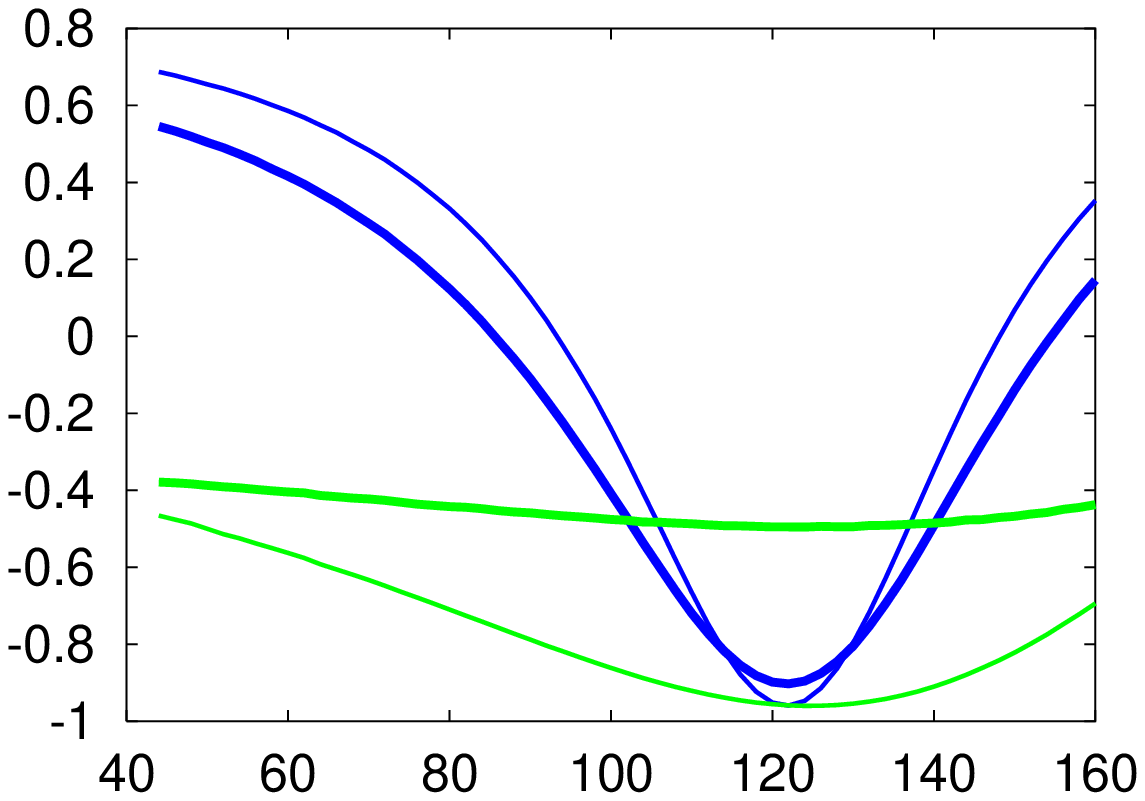}}
\put(5.5,-.3){ \small $M_1${\small /GeV}}
\put(-.2,5.2){\small $ A_{pol}$}
\put(1.9,4.4){\small $(-+),(+-)$}
\put(.9,3.2){\small $(-0),(+-)$}
\put(1.2,2.2){\small $(00),(+-)$}
\put(1,1){\small $(-+),(+-)$}
\end{picture}\par\vspace{.6cm}
\end{minipage}\hfill\hspace{.2cm}
\begin{minipage}{7cm}
\vspace{-1cm}
{\parbox{6.5cm}{\small Figure 10: 
Polarization asymmetries $A_{pol}$, eq.(\re{A_pol}), of
$e^+ e^-\to \tilde{\chi}^0_1 \tilde{\chi}^0_2$, 
$\tilde{\chi}^0_2\to \tilde{\chi}^0_1 e^+ e^-$ at 
$\sqrt{s}=m_{\tilde{\chi}^0_1}+m_{\tilde{\chi}^0_2}+30$~GeV as a function of 
the gaugino parameter $M_1$ for 
$m_{\tilde{e}_L}=176$~GeV, $m_{\tilde{e}_R}=161$~GeV (dark) 
and 
$m_{\tilde{e}_L}=500$~GeV, $m_{\tilde{e}_R}=161$~GeV (light);
the other SUSY parameters as in scenario~A1. 
$(-+)$, $(+-)$ corresponds to
$A_{pol}=(\sigma_e^{-+}-\sigma_e^{+-})/(\sigma_e^{-+}+\sigma_e^{+-})$ and 
analogously for the other curves.}}
\end{minipage}
\end{figure}

\end{document}